\documentclass[pdftex,epjc3]{svjour3}
\usepackage{latexsym}
\usepackage{amssymb}
\usepackage{graphicx,epsfig}
\usepackage{relsize}
\usepackage{graphicx}
\usepackage{amsfonts}
\usepackage{amsmath}
\usepackage{lipsum}
\usepackage{float}
\usepackage{booktabs}
\usepackage{enumerate}
\RequirePackage[colorlinks,citecolor=blue,urlcolor=blue,linkcolor=blue]{hyperref}
\makeatletter
\renewcommand{\footnotesep}{3.5mm}

\newcommand{\singlespacing}{\let\CS=\@currsize\renewcommand{\baselinestretch}{1}\tiny\CS}
\newcommand{\oneandahalfspacing}{\let\CS=\@currsize\renewcommand{\baselinestretch}{1.25}\tiny\CS}
\newcommand{\doublespacing}{\let\CS=\@currsize\renewcommand{\baselinestretch}{1.35}\tiny\CS}

\def\@citex[#1]#2{\if@filesw\immediate\write\@auxout{\string\citation{#2}}\fi
	\def\@citea{}\@cite{\@for\@citeb:=#2\do
		{\@citea\def\@citea{,\linebreak[0]\hskip0pt plus .2em}%
			\@ifundefined{b@\@citeb}%
			{{\bf ?}\@warning{Citation `\@citeb' on page \thepage\space undefined}}%
			\hbox{\csname b@\@citeb\endcsname}}}{#1}}

\date{}
\RequirePackage{graphicx}
\usepackage{float}
\RequirePackage[numbers,sort&compress]{natbib}
\RequirePackage[colorlinks,citecolor=blue,urlcolor=blue,linkcolor=blue]{hyperref}

\begin{document}

\title{Behavior of anisotropic fluids with Chaplygin equation of state in Buchdahl spacetime}
	
	
	\author{Amit Kumar Prasad\thanksref{e1,addr1}
		\and
		Jitendra Kumar\thanksref{e2,addr1} %etc.
		\and
		 Abhijit Sarkar\thanksref{e3,addr1} %etc.
	}
	
	%\thankstext[$\star$]{t1}{Thanks to the title}
	\thankstext{e1}{e-mail:amitkarun5@gmail.com}
	\thankstext{e2}{e-mail:jitendark@gmail.com}
	\thankstext{e3}{e-mail:goovgo@gmail.com}

	\institute{Department of Mathematics, Central University of Jharkhand, Ranchi-835205,  India.\label{addr1}
	}
	
%	\date{Received: date / Accepted: date}

	\maketitle

	\section*{Abstract}
In the present study we have proposed a new model of an anisotropic compact star which admits the Chaplygin equation of state. For this purpose, we consider Buchdahl ansatz. We obtain the solution of proposed model in closed form which is non-singular, regular and well-behaved. In addition to this, we show that the model satisfies all the energy conditions and maintains the hydrostatic equilibrium equation. This model represents compact stars like PSR B0943+10, Her X-1 and SAX J1808.4-3658 to a very good approximate. \\
	\hspace{0.1cm}\\
	\textbf{Key words:} Schwarzschild  Metric;Compact stars; General Relativity.
\section{Introduction}
\subsection{Overview}
Mathematical studies related to the behavior of cosmic bodies, especially compact stars by finding exact solutions of Einstein's field equations (EFE) has been of tremendous importance in the field of theoretical astrophysics and astronomy since the first feather in the cap of general theory of relativity (GR) was added by Schwarzschild's solution \cite{schwarzschild} . The impetus for these studies were renewed after the discovery of the first pulsar by Jocelyn bell in 1967. These compact objects act as lighthouses in the field of observational astrophysics providing unparalleled test-beds for verifying mathematical models. An excellent account of this fact can be found in \cite{Delgaty}. These models also provide interesting insights about the internal geometry and behavior of the nuclear matter since estimating maximum mass-radius ratio depend on the model parameters as well as equation of states (EoS) while considering isotropic as well as anisotropic fluid distributions. It is analyzed in \cite{Sharma} through an investigation of well studied Vaidya-Tikekar model revealing that studies of EoS along with ansatz are equally important for physically acceptable description of super-dense compact objects. The general methodology for investigating the physical behavior and stability criteria of compact stars includes both numerical and analytical approaches taking static, spherically symmetric isotropic or anisotropic solutions preserving hydrostatic equilibrium.
\subsection{Modeling anistropic compact stars}
Class of compact stars is a panorama of stellar formations including not only white dwarfs, neutron stars and black holes with observational support but also hypothetical exotic stars with similar high mass to radius ratio and with extreme nuclear conditions such as high density and temperature. Mathematical models of this cosmic class meditate on a spherically symmetric, self-gravitating and isotropic super-dense descriptions. However, in \cite{Ruderman} Ruderman showed that the interior structure may show anistropic behavior in the density domain of compact stellar objects like neutron stars, boson stars, gravastars etc which is in the order of $10^{15}\,g\,cm^{-3}$,  contrary to the longstanding imposition of isotropic condition of equal radial ($p_r$) and tangential ($p_t$) pressures to solve EFEs. In 1974, with the work of Bowers and Liang \cite{Bowers} anistropic modelings of compact stars were unfolded. These solutions of EFEs considered a vast range of sources for anistropy \cite{Ivanov,Santos,Kippenhahn,Sokolov,Sawyer} along with bringing anisotropic factor, $\Delta = p_t-p_r$ into effect. Also modification of the prior models in isotropic cases became a source for modeling anistropic compact stars \cite{Krori,Maurya3,Prasad,Maurya4,Paul}. Furthermore, incorporating charged relativistic matter obeying different conditions with anisotropic factor \cite{Maurya5,Prasad1,Prasad2,Thirukkanesh,Gomez} helped reveal new breadth of view about the stellar core and subsequently started another important class of endeavor in the field of astrophysical studies combining anisotropy and electromagnetic fluid spheres. Analytical approaches to construct these models use different techniques to find solutions, for example using different metric potentials, imposing symmetry constraints, applying different equation of states or combinations of these. Evidently, newer ideas and methods for modeling anisotropic compact stars are expanding the horizon of these studies to encompass modeling in the realm of modified gravitational theories along with exotic, yet-to-discover stellar objects.
%%%%%%%%%%%%%%%%%%%%%%%%%%
\subsection{EoS method}
To describe the interior structure of the stellar system it is needed to apply methods for solution to tackle the hardship of solving highly non-linear EFEs. Assumption of relation between the radial pressure and energy density, $p = p(\rho)$, is known as the equation of state (EOS) method. Evolution of this method has seen implementation of linear, quadratic, polytropic, Chaplygin and various other relations to avail different anisotropic models \cite{Singh,Maharaj,Ngubelanga,Buchdahl,Herrera1,Rahaman1}. Discernibly, the significance of EoS method in modeling a compact stellar object is reflected in the analysis of physical behavior. This has stimulated researchers to consider subtler yet more generalized forms of EoSs as well as proposing new physical nature of the anistropic fluid matter \cite{Fuenmayor,Lobo,Azam,Maurya4,Maedan,Otto,Koliogiannis,Ortiz}. Hence this method is crucial not only for modeling already discovered interesting stellar bodies but to propose feasible answers to unsolved puzzles and theoretical conundrums in the field of relativistic astrophysics and cosmology.
\subsection{Importance of Chaplygin EoS}
The EoS of Chaplygin gas model is a particular form of polytropic EOS described as $p =  \dfrac{\textsl{C}}{\rho^{\alpha}}$ where $p$,$\rho$ are pressure and density with $\textsl{C}$ being a positive constant and the parameter $\alpha = 1$ or in generalized EoS $0<\alpha \le 1$. This form of the EoS gives it an edge to model exotic stellar objects like the class of compact stars (See \cite{Bento,kamenshchik}) withstanding extreme physical conditions. Since the stability in the interior of compact objects is essentially maintained by halting the gravitational collapse of the stellar mass by various conditions such as hydrostatic force, Coulomb's force from the presence of electric charge etc, the generalized Chaplygin gas EOS is very useful in order to incorporate new theories like dark fluid model \cite{Farnes} or dark energy stars \cite{Rahaman1} as another repulsive force adding stability to the compact star model.
\subsection{Our work}
In this paper, we derived a new model for anisotropic compact stars with the modified Chaplygin
equation of state described by $p_r = \beta_1\,\rho + \dfrac{\beta_2}{\rho}$ , coupled with a suitable form of gravitational potential namely Buchdahl ansatz. We have showed that our solution is stable, physically feasible, nonsingular, continuous and maintains hydrostatic equilibrium in the interior of the star. The workflow of our present article is as follows: Solutions of Einstein's field equations and model parameters have been discussed with graphical analysis of the solution in Sec.\ref{Sec2} and Sec.\ref{Sec3} with respect to a particular form of Chaplygin EoS along with Buchdahl ansatz. Sec.\ref{Sec4} describes the required conditions for interior metric to join the exterior one smoothly. A detailed discussions on physical feasibility, stability of our model and influence of choice of the EoS in the physical behavior of the compact stars are presented in Sec.\ref{Sec5}. In Sec.\ref{Sec6} we briefed our work and concluded with notable remarks.
\section{Field equation} \label{Sec2}
	Let us consider the static spherically symmetric metric in Schwarzschild coordinates
	
\begin{eqnarray}
		ds^{2}=-e^{\lambda(r)}dr^{2} -r^{2}(d\theta^{2}+\sin^{2}\theta d\phi^{2})+e^{\nu(r)}dt^{2}\label{1}
\end{eqnarray} 

where $\lambda(r)$ and $\nu(r)$ are the functions of radial coordinate $r.$  If (\ref {1}) describes anisotropic matter distribution then the space-time(\ref {1}) has to satisfy the energy-momentum tensor

\begin{eqnarray}  T^{i}_{j} = (\rho+p_t)u^{i}u_{j}-p_t\delta^{i}_{j}+(p_r-p_t)\chi^{i}\chi_{j}\label{2}
\end{eqnarray}

with $ u^{i}u_{j}= - \chi^{i}\chi_{j} = 1 $, where the vector $u^{i}$ is the fluid four velocity and  $\chi_{i}$ is the unit space-like vector which is orthogonal to $u^{i}$, i.e. $ u^{i}\chi_{i} = 0 $. However $\rho,\,p_r, $ and  $p_t$ represent the density, radial pressure and tangential  pressure respectively. Thus, the Einstein field equation for line element (\ref{1}) with respect to energy-momentum tensor (\ref{2}) reduce to following equations in terms of physical parameters and metric function as    (suppose $ G = c = 1 $)
\begin{eqnarray}
		8\pi p_r &=&\dfrac{\nu'}{r}e^{-\lambda}-\dfrac{(1-e^{-\lambda})}{r^{2}}\label{3} \\
		8\pi p_t&=&\bigg(\dfrac{\nu''}{2}-\dfrac{\lambda'\nu'}{4}+\dfrac{\nu^{'2}}{4}+\dfrac{\nu'-\lambda'}{2r}\bigg)e^{-\lambda}\label{4}\\
		8\pi\rho&=&\dfrac{\lambda'}{r}e^{-\lambda}+\dfrac{(1-e^{-\lambda})}{r^{2}}\label{5}
\end{eqnarray}
where ($ ' $) prime denotes the differentiation with respect to radial coordinate $r$. Using the eqs. (\ref{3}) and (\ref{4}), we get

\begin{equation}
		\Delta = 8\pi \, (p_{t} -\, p_{r} ) = e^{-\lambda } \left[\frac{\nu''}{2} -\frac{\lambda' \nu'}{4} +\frac{{\nu'}^{2} }{4} -\frac{\nu'+\lambda '}{2r} -\frac{1}{r^{2} } \right]\,  +\frac{1}{r^{2} }. \label{6}
\end{equation}
	here $ \Delta = 8\pi \, (p_{t} -\, p_{r} ) $ is denoted as the
	anisotropy factor and it measures the pressure anisotropy of the fluid. The anisotropic pressure is repulsive if $ p_t > p_r $ and attractive if $ p_t < p_r $ of the stellar model. 
	
	The assumption of the equation of state is very important to derive a physically motivated stellar model, which relates the pressure to the density of the star, i.e., $ p = p(\rho) $. Most of the earlier works were centered on imposing a linear equation of state of the form $ p = \alpha\rho $ where $ \alpha $ is a constant. Here we use the Chaplygin equation of state (EoS) of the form 
\begin{eqnarray} 
		p_r = \beta_1\,\rho + \dfrac{\beta_2}{\rho}\label{7}
\end{eqnarray}
	for the solving the system of equations, where $ \beta_1 $ and $ \beta_2 $ are positive constants. Rahaman et al.\cite{Rahaman}, Bhar \cite{Bhar} and Benaoum \cite{Benaoum} have used the Chaplygin equation of state (EoS) to model compact stars within the framework of general relativity.
\section{Solution of the model for anisotropic stars} \label{Sec3}
	To solve the Einstein field equations we consider a known metric Buchdahl ansatz\cite{Buchdahl}  which is given by
\begin{eqnarray} e^{\lambda}=\frac{K\,(1+Cr^2)}{K+Cr^2}, ~~~~\textrm{with}~~ K<0 ~~~ \textrm{and} ~~~ K>1\label{8} 
\end{eqnarray} 
where $ K $ and $ C $ are constant parameters that characterize the geometry of the star. Initially the Buchdahl\cite{Buchdahl} have considered above metric potential to study a relativistic compact star. Note that above metric potential is free from singularity at $ r = 0 $ and the metric coefficient is $ e^{\lambda(0)} = 1.$ Here we pull out the range of $ 0 < K < 1 $, because in this range either the energy density or pressure will be negative depending on the two parameters. In a
more generic situation, we gain Vaidya-Tikekar\cite{Vaidya} metric when $ C=-K/R^2 $. In our analysis we introduce the transformation $ x= r^2,~~  e^{\nu} = y^2(x)$  and substituting the value of $e^{\lambda} $ into the eqs. (\ref{3})-(\ref{5}), we get
\begin{eqnarray}
		8\pi\,\rho &=& \dfrac{C\,(K-1)(3+Cx)}{K(1+Cx)^2}\label{9} \\
		8\pi\,p_r &=& 4\bigg( \frac{K+Cx}{K(1+Cx)} \bigg)\dfrac{\dot{y}}{y} + \dfrac{C\,(1-K)}{K(1+Cx)} \label{10}\\
		p_t &=& p_r + \Delta \label{11}
\end{eqnarray}  
	where a dot denotes the derivative with respect to x and the expression of  $ \Delta $ is 
\begin{eqnarray}
		8\pi\,\Delta =  4x\bigg( \frac{K+Cx}{K(1+Cx)} \bigg)\dfrac{\ddot{y}}{y} - \bigg(\dfrac{2Cx(K-1)}{K(1+Cx)^2}\bigg)\dfrac{\dot{y}}{y} + \dfrac{C^2x(K-1)}{K(1+Cx)^2}    \label{12}
\end{eqnarray}

Substituting the eqs.(\ref{9}) and (\ref{10}) into the eq.(\ref{7}) we get
\begin{eqnarray}
		\dfrac{\dot{y}}{y} = \dfrac{\beta_1\,C(K-1)(3+Cx)}{4(K+Cx)(1+Cx)} + \dfrac{64\pi^2\,\beta_2\,K^2(1+Cx)^3}{4C(K-1)(K+Cx)(3+Cx)} + \dfrac{C\,(K-1)}{4(K+Cx)}  \label{13}
\end{eqnarray}

Integrating the eq.(\ref{13}) we obtain

\begin{equation}
\begin{aligned}
			\log(y)  = \dfrac{\beta_1\,(K-1)}{4\,(1-K)}\bigg[\,(3-K)\log(K+Cx) -8\,\log(1+Cx) \,\bigg] + \dfrac{\beta_2\,64\pi^2\,K^2}{4\,C(K-1)}\bigg[ \dfrac{Cx^2}{2} - Kx \\
			- \dfrac{8\log(3+Cx)}{C(K-3)} + \dfrac{(K-1)^3\,\log(K+Cx)}{C(K-3)} \bigg] + \dfrac{(K-1)\log(K+Cx)}{4} + D \label{14}
\end{aligned}
\end{equation}
where $ D $ is arbitrary constant of integration. Subsequently, density, the radial pressure, tangential pressure and anisotropic factor ($ \Delta $) are obtained as 

\begin{eqnarray}
\rho &=& \dfrac{C\,(K-1)(3+Cx)}{8\pi\,K(1+Cx)^2} \label{15}\\
p_r &=& \dfrac{\beta_1\,C\,(K-1)(3+Cx)}{8\pi\,K(1+Cx)^2} + \dfrac{\beta_2\,8\pi\,K(1+Cx)^2}{C\,(K-1)(3+Cx)}  \label{16}\\
p_t &=& p_r + \Delta \label{17}
\end{eqnarray}
\begin{eqnarray}
		\Delta &=&  \dfrac{\beta_1\,C^2\,x(1-K)}{16\pi\,K(K+Cx)(1+Cx)^3}\bigg[ 3 + 7K + 11Cx + KCx + 2\,C^2x^2 \bigg] \nonumber\\
		&&+\dfrac{4\beta_2\,\pi\,Kx(1+Cx)}{(K-1)(K+Cx)(3+Cx)^2}\bigg[  -3+13K+9\,Cx+3KCx+2\,C^2x^2\bigg]\nonumber \\
		&& -\dfrac{C^2x(K-3)(K-1)^2}{16\pi\,K(K+Cx)(1+Cx)^2} + 4x\,\big(F(x)\big)^2 \label{18}
\end{eqnarray}
	
where $ F(x) = \dfrac{C(K-1)\Big( \beta_1\,(3+Cx)+1+Cx\Big)}{4(K+Cx)(1+Cx)} + \dfrac{64\pi^2\,\beta_2\,K^2(1+Cx)^3}{4C(K-1)(K+Cx)(3+Cx)}  $ \\

\begin{figure}[H]
\begin{center}
\includegraphics[width=6cm]{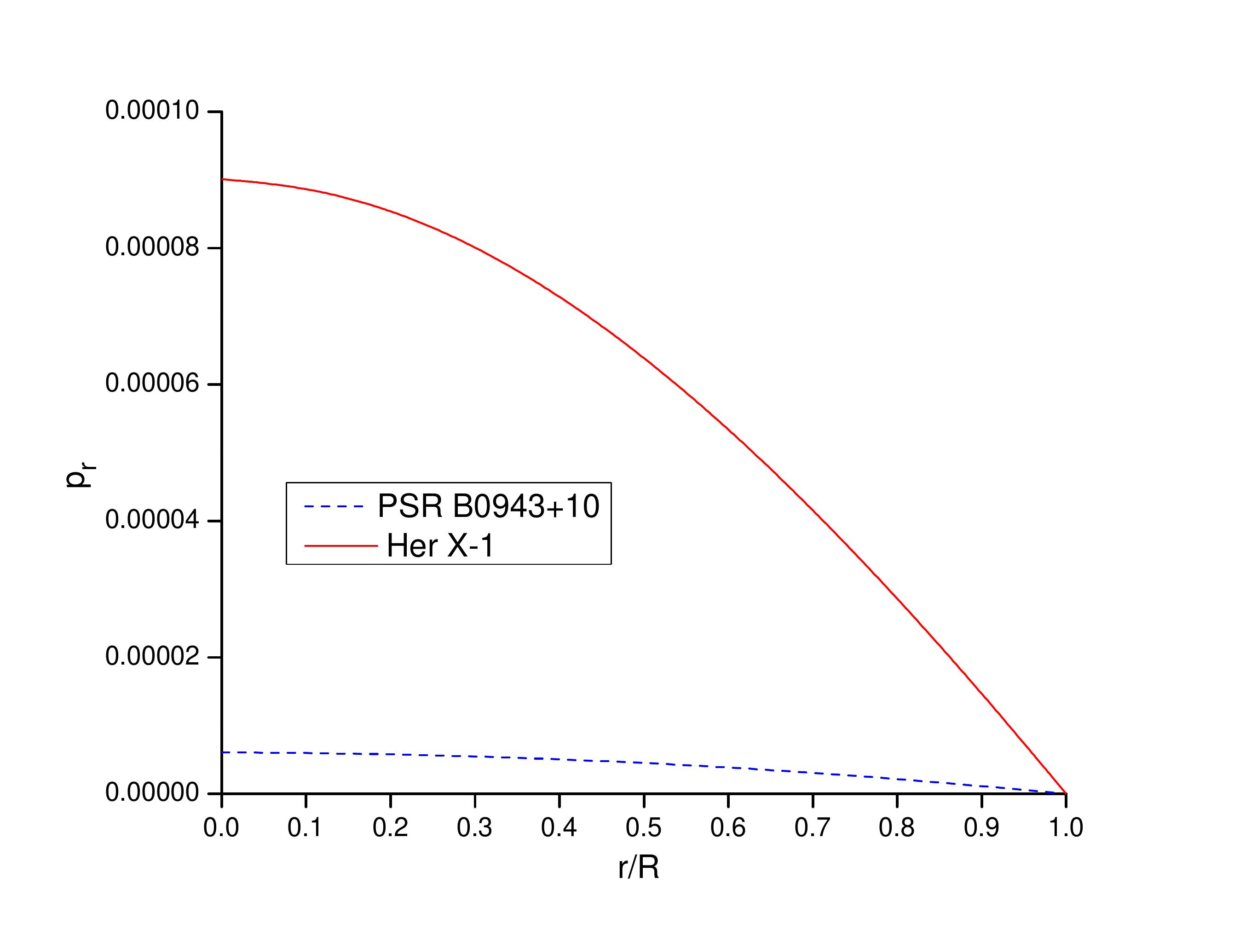}
\includegraphics[width=6cm]{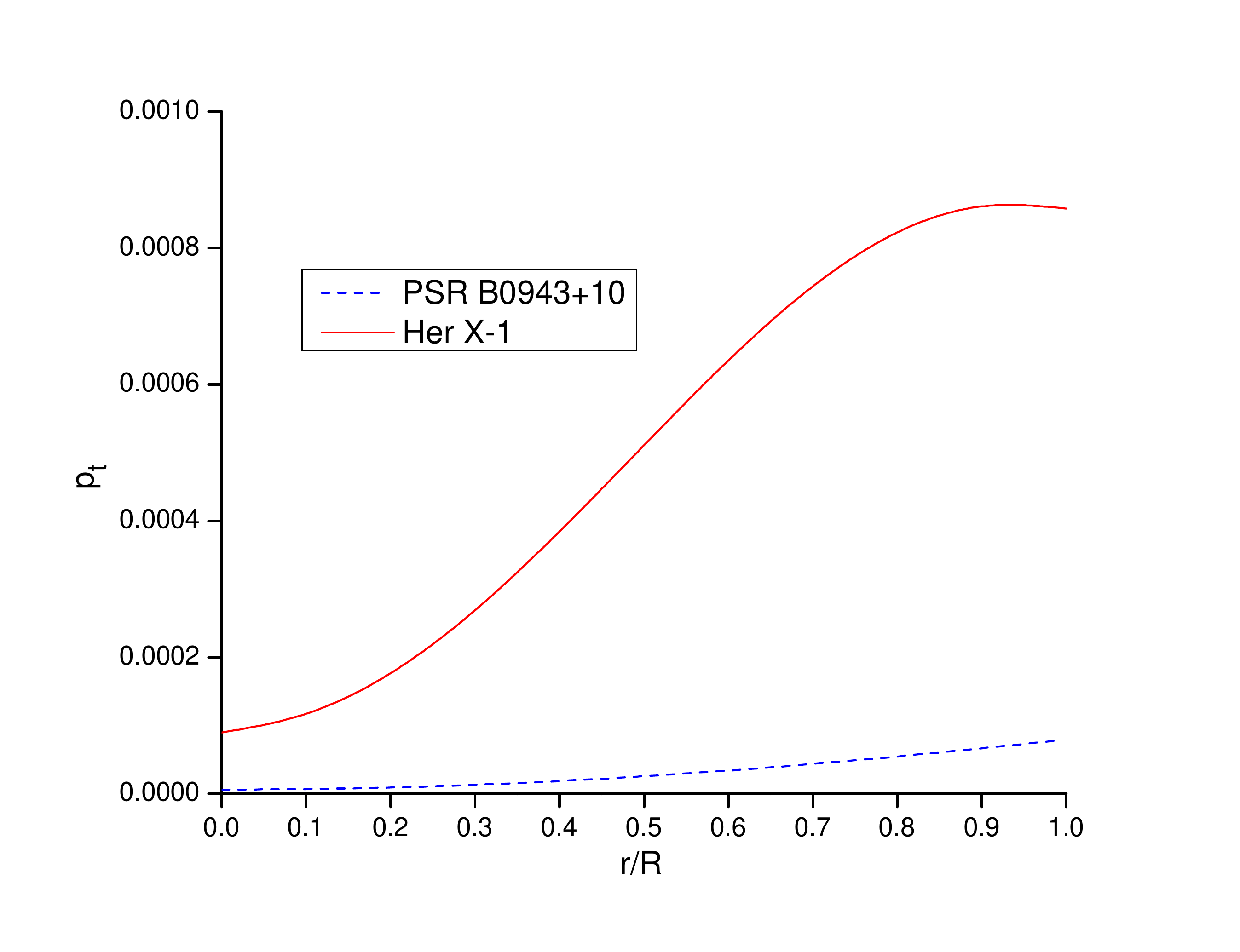}\\
\includegraphics[width=6cm]{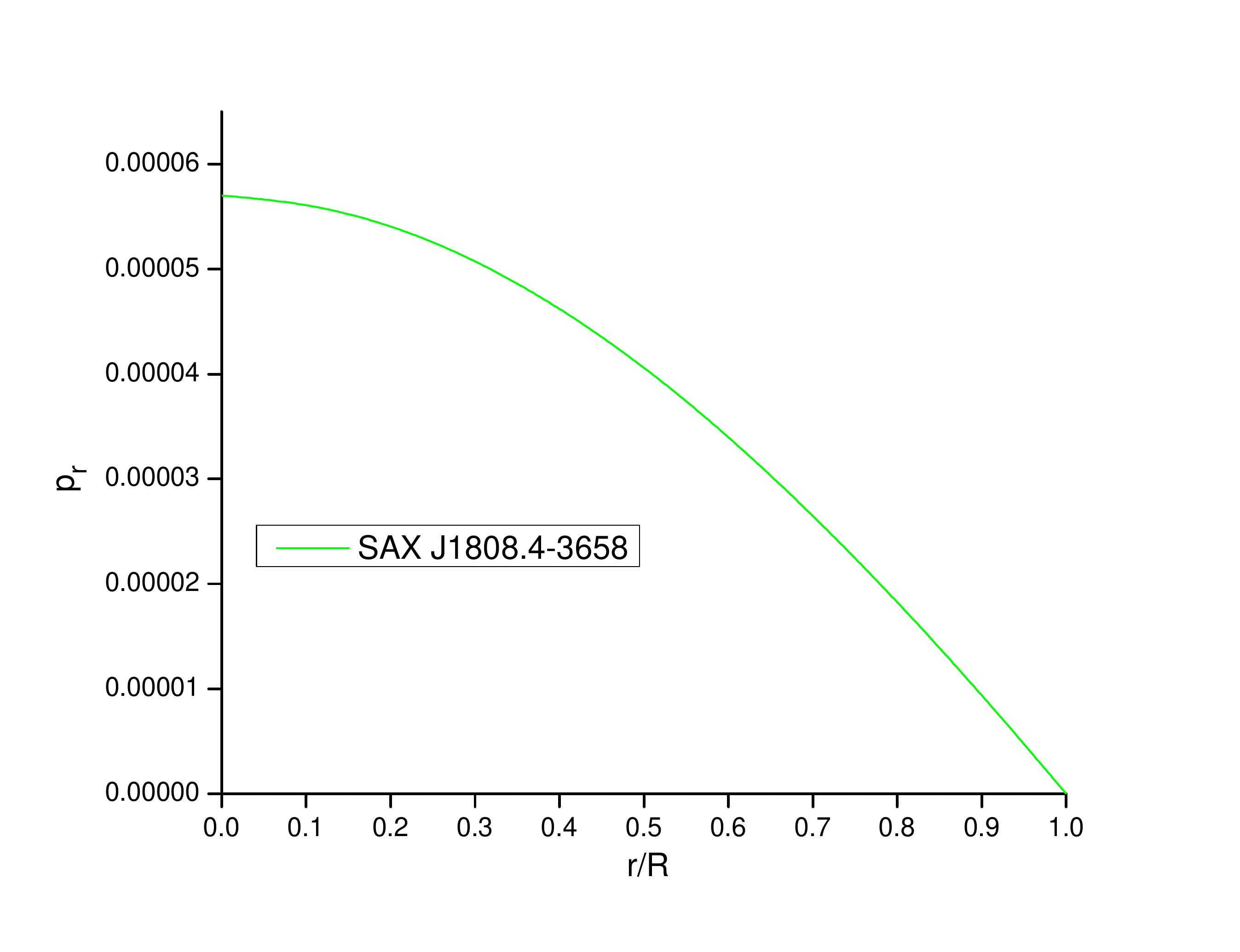}
\includegraphics[width=6cm]{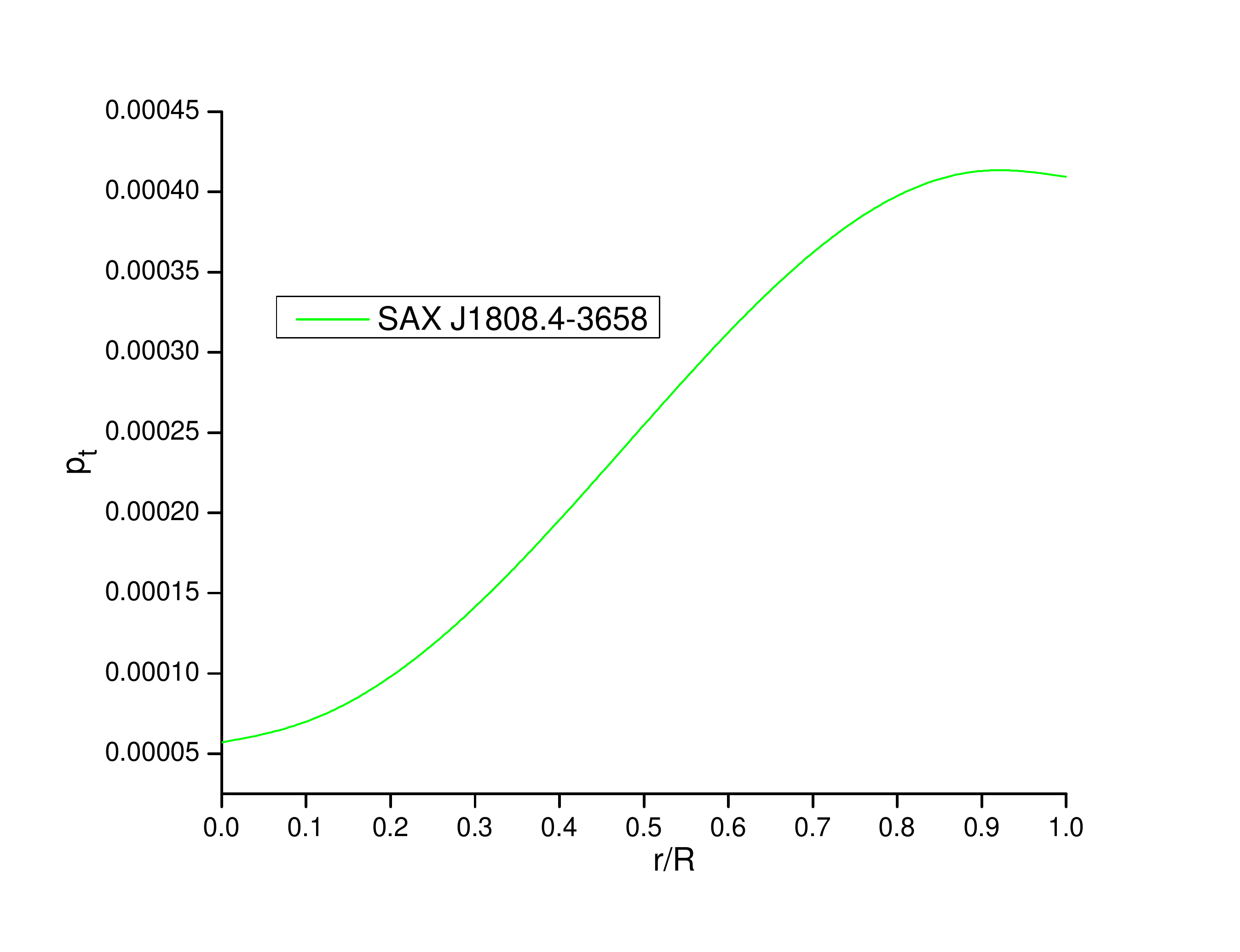}
\caption{ The radial pressure($ p_r $) and transverse pressure ($ p_t $) against radial coordinate $ r/R $ have been plotted for $ K<0 $ (top) and $ K>1 $ (bottom). The values of the parameter which we have used for graphical presentation are:(i) $ K=-1.52,\,\,\beta_1 = 0.32, \,\, C=0.00211km^2,\,$ $ M=0.02M_{\odot},\,\, R=2.575km  $, for PSR B0943+10; (ii) $ K=-11.20,\,\,\beta_1 = 0.3, \,\, C=0.00408km^2,\,$ $ M=0.07M_{\odot},\,\, R=8.43km $ for Her X-1 and (iii) $ K=14.5,\,\,\beta_1 = 0.28, \,\, C=0.00337km^2,\,$ $ M=0.06M_{\odot},\,\, R=8.951km  $ for SAX J1808.4-3658. See Tables-\ref{T1} and \ref{T2} for more details.}\label{f1}
\end{center}
\end{figure}
\begin{figure}[h]
\begin{center}	\includegraphics[width=6cm]{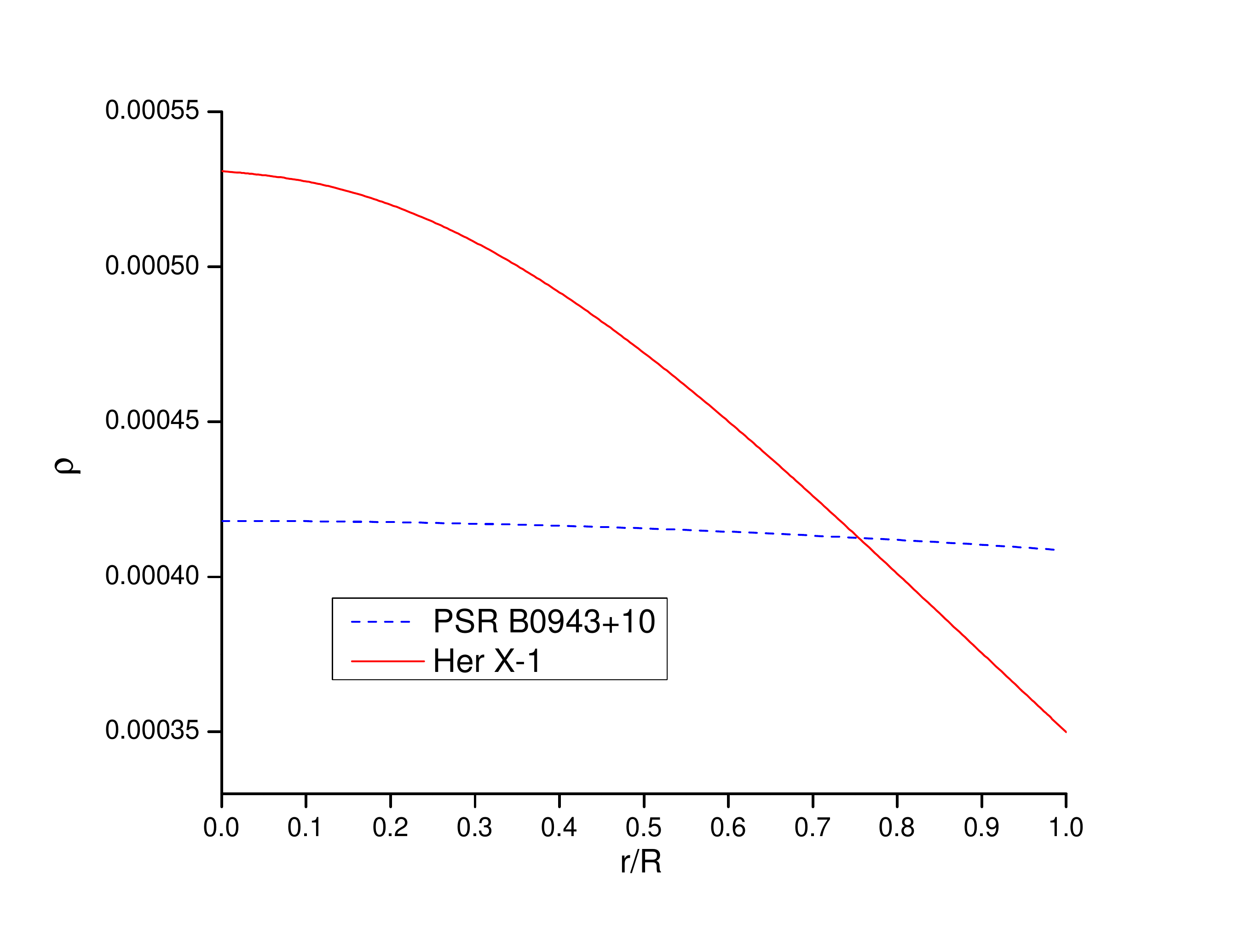}\includegraphics[width=6cm]{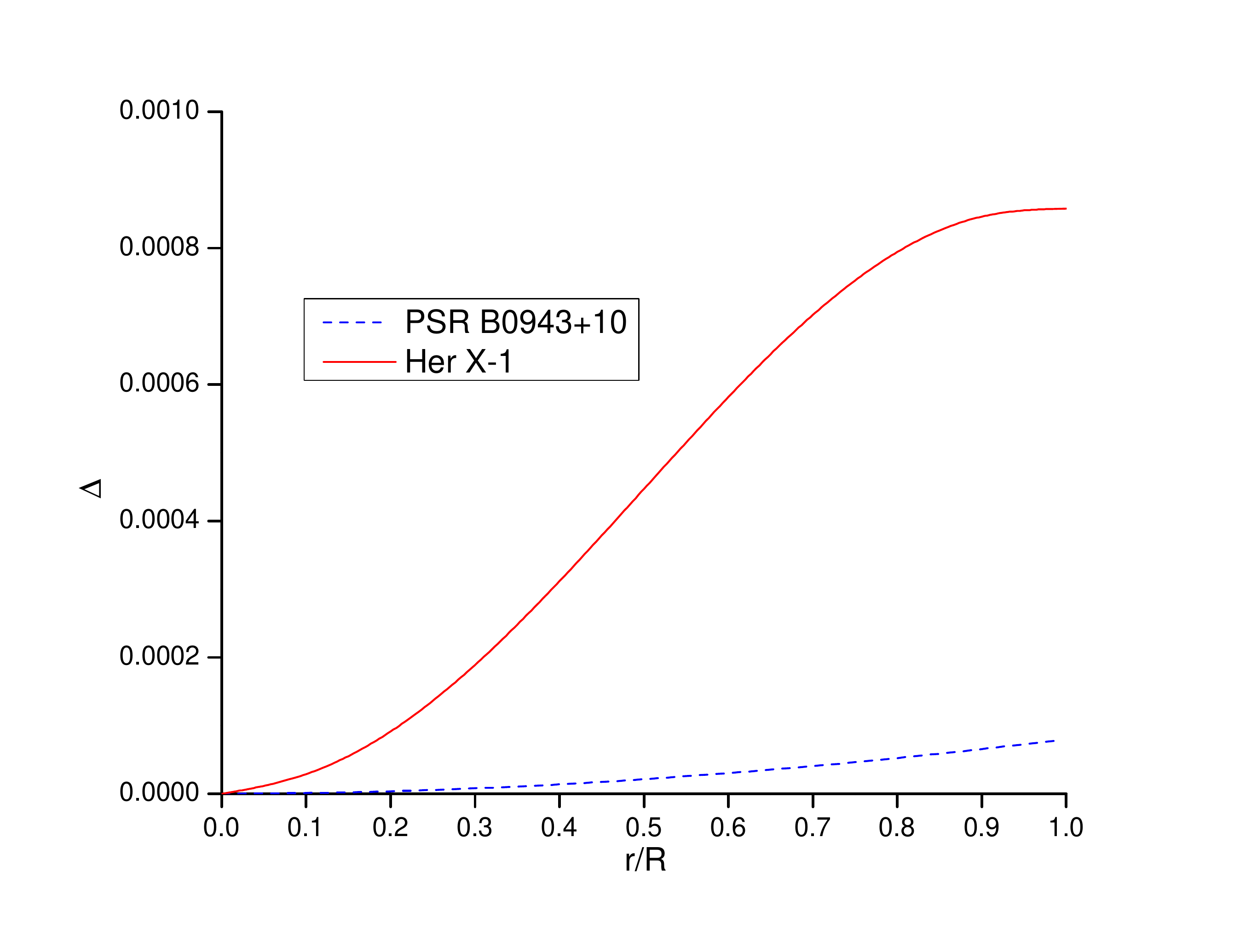}\\\includegraphics[width=6cm]{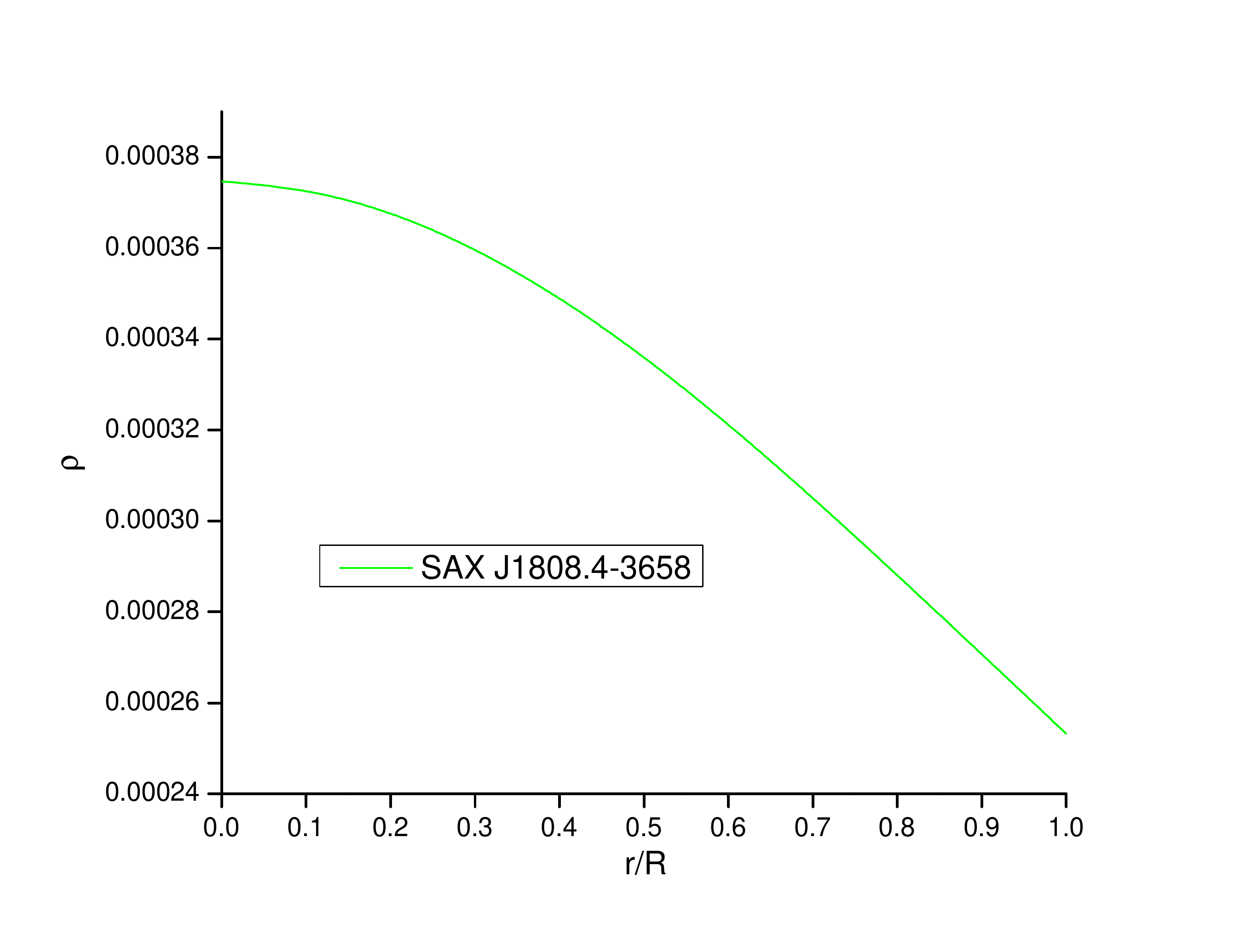}\includegraphics[width=6cm]{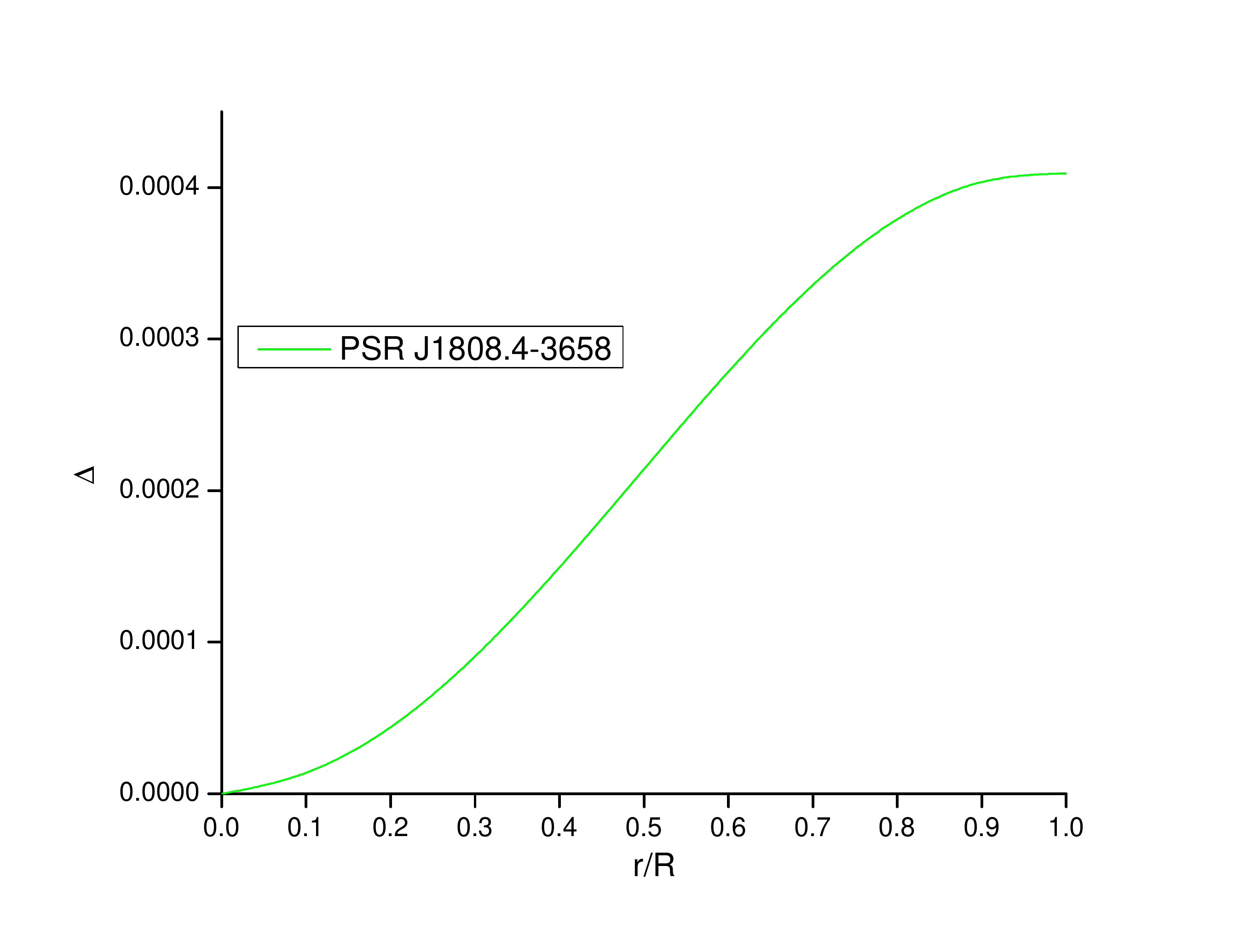}
\caption{Variation of energy density($ \rho $) and anisotropy($ \Delta $) versus radial coordinate $ r/R $ for $ K<0 $ (top) and $ K>1 $(bottom). For plotting these graphs, we have used same data set as Fig.\ref{f1}. }\label{f2}
\end{center}
\end{figure}
	The physical validity of relativistic stellar model depends on some conditions throughout the stellar interior: (a) the density and pressure should be positive definite at the center, (b) the density and pressure should be maximum at center and decreasing monotonically within $ 0 < r < R $. To analyze these features, we plot the Figs.\ref{f1} and \ref{f2} for our stellar model for  $ K<0  $  and $ K>1 $ .  We have observed from these plots that the  pressure and density decrease towards the boundary and maximum at the center. Also the anisotropy is positive in the model and zero at the center, this shows that the tangential pressure ($ p_t $) is always greater than the radial pressure ($ p_r $).
\section{Junction conditions} \label{Sec4}
	In order to smoothly match the interior spacetime metric (\ref{1}) to the vacuum  Schwarzschild exterior solution at the boundary ($ r=R $) which is given by the line element 
	\begin{eqnarray}
		ds^{2}=-\bigg(1-\frac{2M}{r}\bigg)^{-1}dr^{2}-r^{2}(d\theta^{2}+\sin^{2}\theta d\phi^{2})+\bigg( 1-\frac{2M}{r}\bigg)dt^{2} \label{19}
	\end{eqnarray}
	For this purpose we impose the Israel-Darmois junction conditions \cite{Israel,Darmois}. The we have  \begin{eqnarray} e^{-\lambda}&=&1-\dfrac{2M}{R},~~~\textrm{and}~~~e^{\nu}=y^{2}=1-\dfrac{2M}{R} \label{20}\\	 p_{r}(R)&=&0.\label{21}
	\end{eqnarray}
	Using the conditions (\ref{20}) and (\ref{21}),  we obtain the   
	\begin{equation}
		\begin{aligned}
			D =	\dfrac{1}{2}\log\bigg( 1 - \frac{2M}{R} \bigg) - \dfrac{\beta_1\,(K-1)}{4\,(1-K)}\bigg[\,(3-K)\log(K+CR^2) -8\,\log(1+CR^2) \,\bigg] - \dfrac{\beta_2\,64\pi^2\,K^2}{4\,C(K-1)}\\\times \bigg[ \dfrac{CR^4}{2} - Kx 
			- \dfrac{8\log(3+CR^2)}{C(K-3)} + \dfrac{(K-1)^3\,\log(K+CR^2)}{C(K-3)} \bigg] - \dfrac{(K-1)\log(K+CR^2)}{4}  \label{22}
		\end{aligned}
	\end{equation} 
	and 
	\begin{eqnarray}
		\dfrac{\beta_2}{\beta_1} &=& \dfrac{C^2\,(K-1)^2(3+CR^2)}{64\pi^2\,K^2(1+CR^2)^2}\label{23}\\
		M &=& \dfrac{CR^3(K-1)}{2K\,(1+CR^2)}\label{24}
	\end{eqnarray}
	We have demonstrated the values of constant parameters $ C,\, K,\, \beta_1 $ and $ \beta_2 $  in Table-\ref{T1} \& \ref{T2}.
\section{Physical acceptability and stability conditions for anisotropic compact star} \label{Sec5}
	In this section, we have discussed the various physical properties of our solution. We analyzed the stability and acceptability problem through different conditions which is given below as follows:
	\subsection{Equilibrium condition}
\begin{figure}[h]
\begin{center}
	\includegraphics[width=5.5cm]{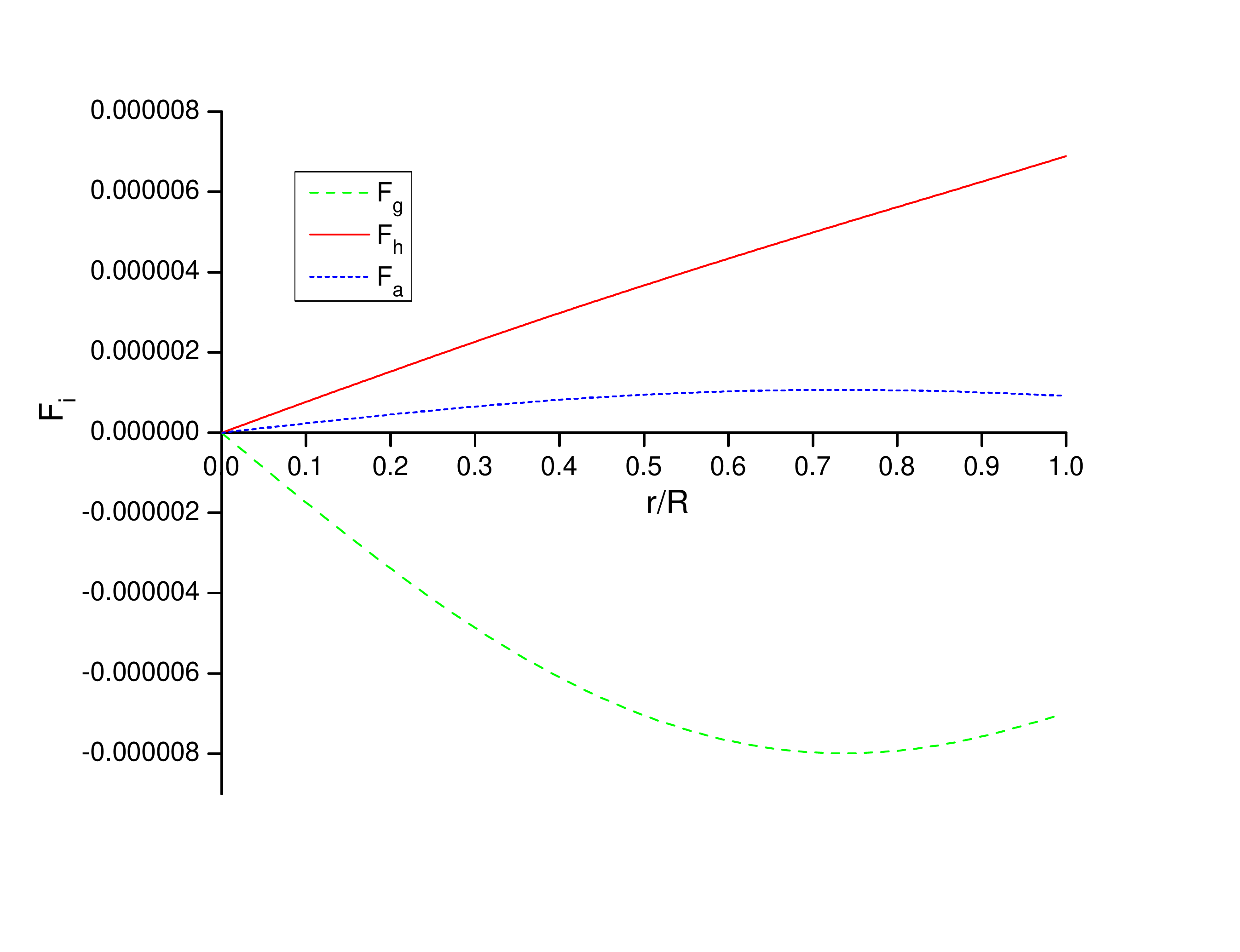}\includegraphics[width=5.5cm]{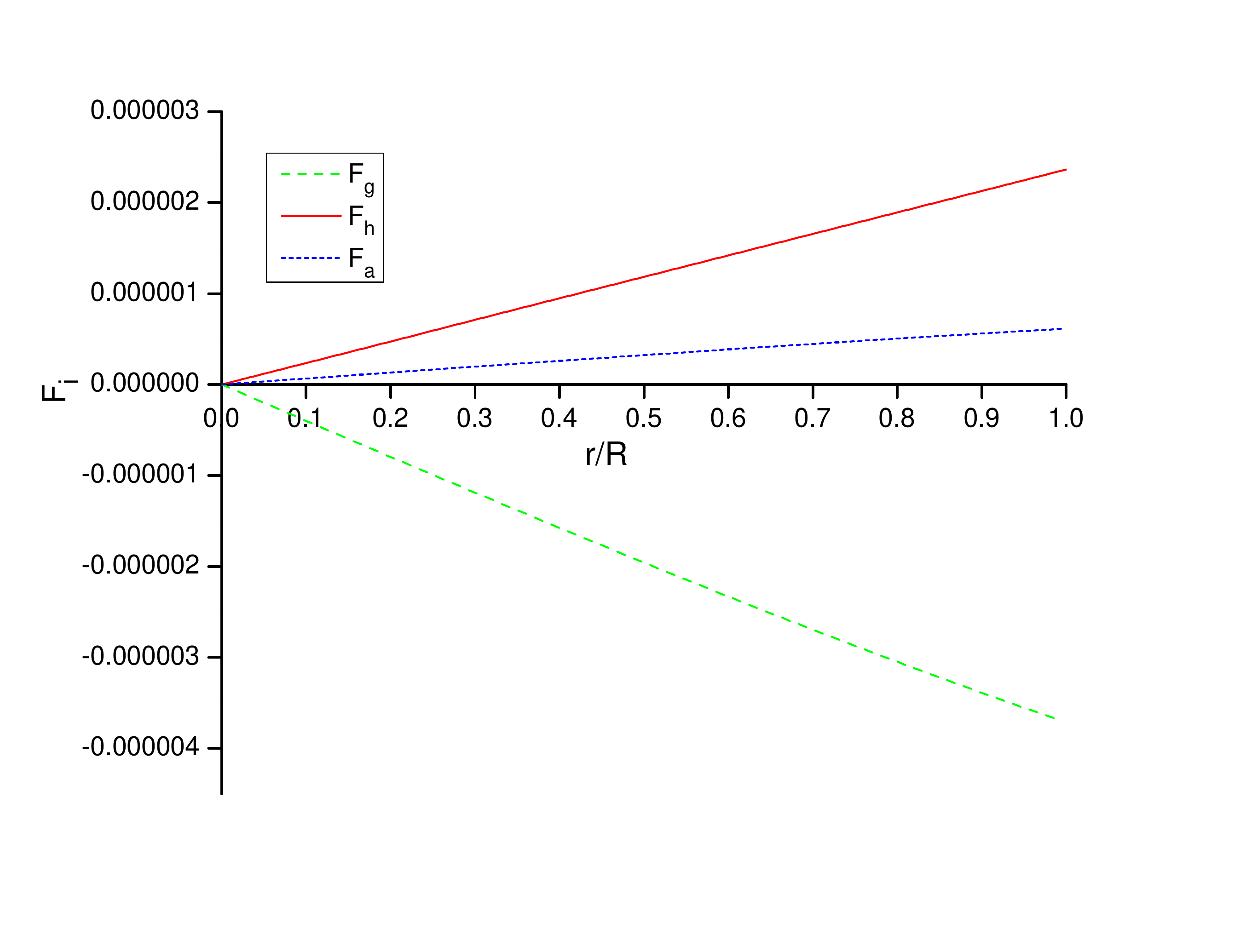}\includegraphics[width=5.5cm]{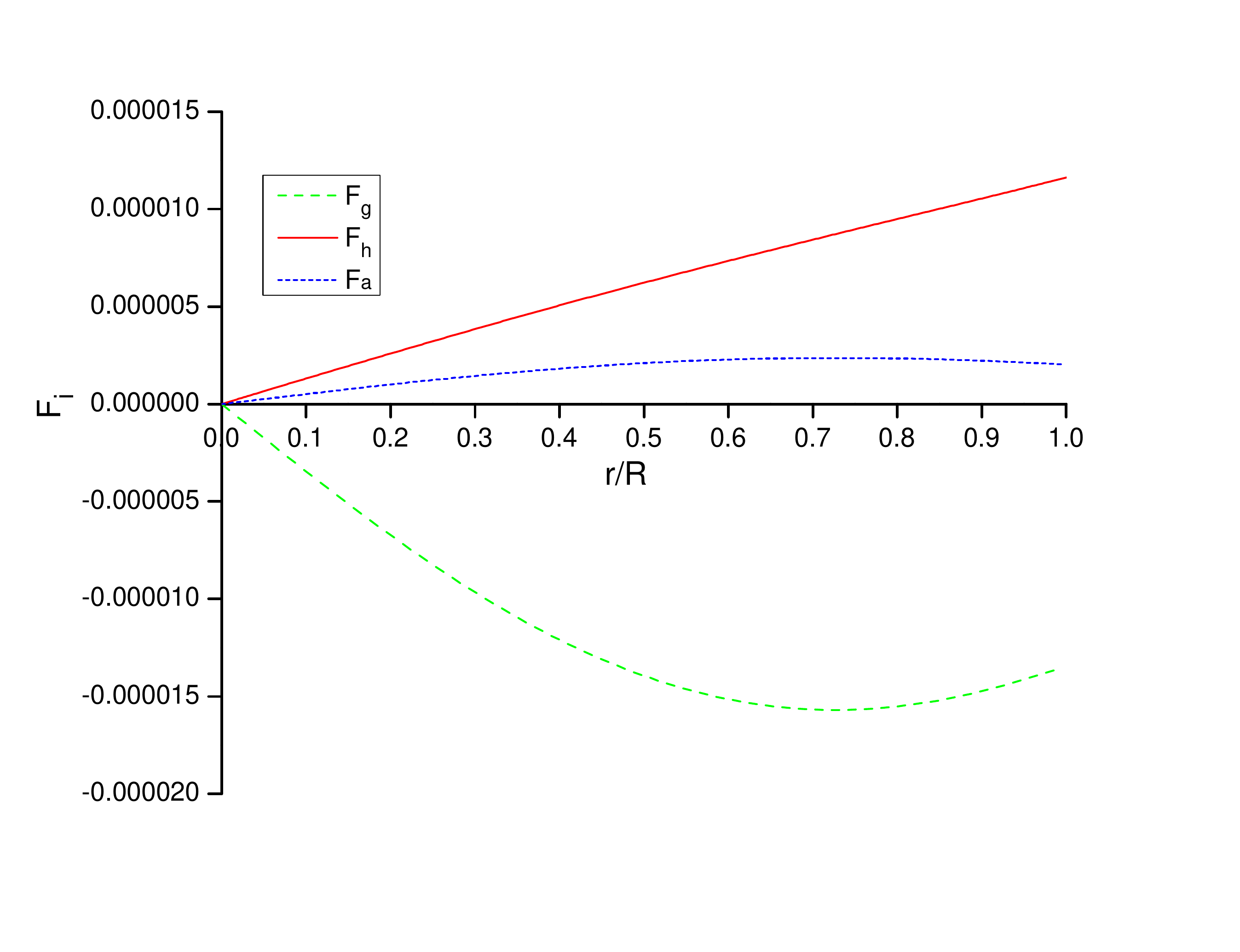}
	\caption{Variation of different forces versus radial coordinate $ r/R $. For this figure we have used the following data  (i) $ K=14.5,\,\,\beta_1 = 0.28, \,\, C=0.00337 km^2,\,$ $ M=0.06M_{\odot},\,\, R=8.951 km  $ for SAX J1808.4-3658 (first graph), (ii) $ K=-1.52,\,\,\beta_1 = 0.32, \,\, C=0.00211 km^2,\,$ $ M=0.02M_{\odot},\,\, R=2.575 km  $, for PSR B0943+10(second graph) and (iii) $ K=-11.20,\,\,\beta_1 = 0.3, \,\, C=0.00408 km^2,\,$ $ M=0.07M_{\odot},\,\, R=8.43 km $ for Her X-1(third graph). }\label{f3}
\end{center}
\end{figure}

	To check the equilibrium condition of our stellar model we have considered the Tolman-Oppenheimer-Volkoff(TOV) equation\cite{35,36} which is given by
	\begin{eqnarray}
		-\frac{M_G(\rho+p_r)}{r^2}e^{\frac{\lambda-\nu}{2}}-\frac{dp_r}{dr}+
		\frac{2\Delta}{r} =0,\label{25} 
	\end{eqnarray}
	where $M_G$ is the effective gravitational mass given by:
	\begin{eqnarray}
		M_G(r)=\frac{1}{2}r^2 \nu^{\prime}e^{(\nu - \lambda)/2}\label{26}
	\end{eqnarray}
	From the equation (\ref{26}) the value of $M_G(r)$ put in the equation (\ref{25}), we get
	\begin{eqnarray}
		-\frac{\nu'}{2}(\rho+p_r)-\frac{dp_r}{dr}+\frac{2\Delta}{r} =0,  \label{27}
	\end{eqnarray}
	The equation (\ref{27}) describes the equilibrium condition for anisotropic stellar model through gravitational force $(F_g)$, hydrostatic force $(F_h)$ and anisotropic force $(F_a)$ with the expressions 
	\begin{eqnarray}
		F_g&=&-\frac{\nu'}{2}(\rho+p)= -\dfrac{y'\,r}{4\pi y}\bigg[ \dfrac{C\,(\beta_1+1)\,(K-1)(3+Cr^2)}{K(1+Cr^2)^2} + \dfrac{\beta_2\,64\pi^2\,K(1+Cr^2)^2}{C\,(K-1)(3+Cr^2)}   \bigg]\label{28}\\
		F_h&=&-\frac{dp}{dr}=-\dfrac{1}{4\pi}\bigg[  \dfrac{\beta_1\,C^2(K-1)^2(3+Cr^2)^2 - 64\beta_2\,\pi^2\,K^2(1+Cr^2)^4}{C^2(K-1)^2(3+Cr^2)^2}\bigg] \nonumber\\&&~~~~~~~~~\times \bigg(\dfrac{C^2r(1-K)(5+Cr^2)}{K(1+Cr^2)^3} \bigg) \label{29}\\
		F_a&=&\frac{2\Delta}{r}\label{30}
	\end{eqnarray}
	Fig.\ref{f3} display the gravitational force $(F_g)$, hydrostatic force $(F_h)$ and anisotropic force $(F_a)$ are regular and finite at center as well as on the surface of compact stars PSR B0943+10, Her X-1 (for $ K<0 $) and SAX J1808.4-3658(for $ K>1 $). We can observe from these figures that the gravitational force $(F_g)$ is counterbalanced by the combined effects of hydrostatic force $(F_h)$ and anisotropic force $(F_a)$. 
\begin{table}[h]
	\begin{center}
		\caption{ Values of the masses $ M $, radii $ R $, and the constants $ C $, $ K $, and $ \beta_2 $ for the compact stars.}\label{T1}
		\begin{tabular}{lllllll}
			\hline
			Compact star&  Mass & Observed  &  Predicted  & $ C  $ & $ K $&$ \beta_2 $ \\
			Candidates	& $ M(M_{\odot}) $ &Radius $ R(km) $ & Radius $R(km) $ & $ km^{-2} $ & & $ km^{-4} $ \\
			\hline
			PSR B0943+10 (Yue et al.\cite{Yue})& 0.02 & 2.6 & 2.575 & 0.00211 & -1.52&  $ -5.339\times10^{-8} $\\
			\hline
			Her X-1 (Abubekerov et al.\cite{Abubekerov})& $ 0.85\pm 0.15 $ & $ 8.1\pm 0.41 $ & 8.43 & 0.00408 &  -11.20 & $ -3.672\times10^{-8} $\\
			\hline
			SAX J1808.4-3658 (Elebert et al.\cite{Elebert})	& $ 0.9\pm 0.3 $ & $ 7.951\pm 1.0 $ & 8.951 & 0.00337 & 14.5 & $ -1.795\times10^{-8} $ \\
			\hline
		\end{tabular}	
	\end{center}
\end{table}

\subsection{Causality condition}
\begin{figure}[h]
\begin{center}
\includegraphics[width=5.5cm]{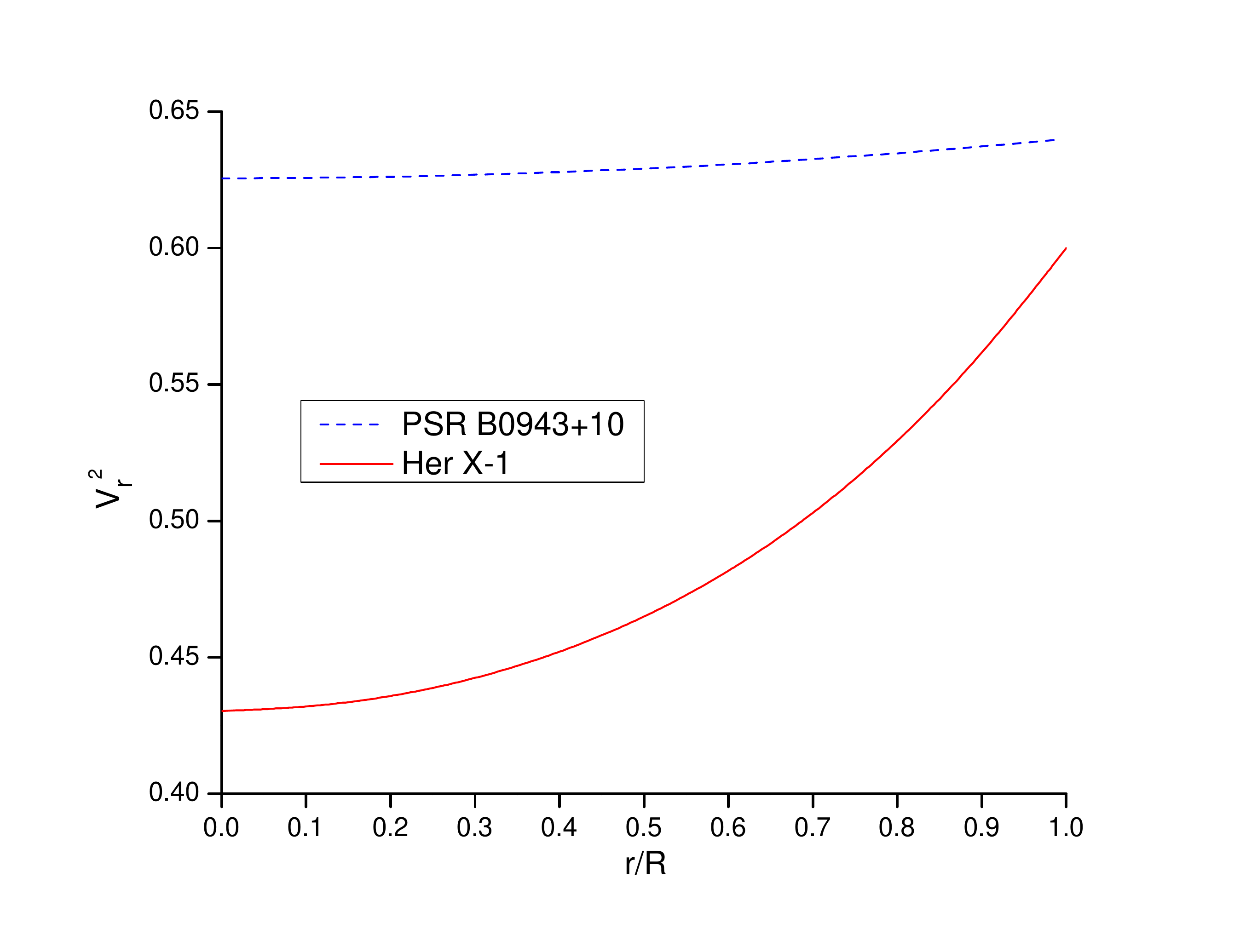}\includegraphics[width=5.5cm]{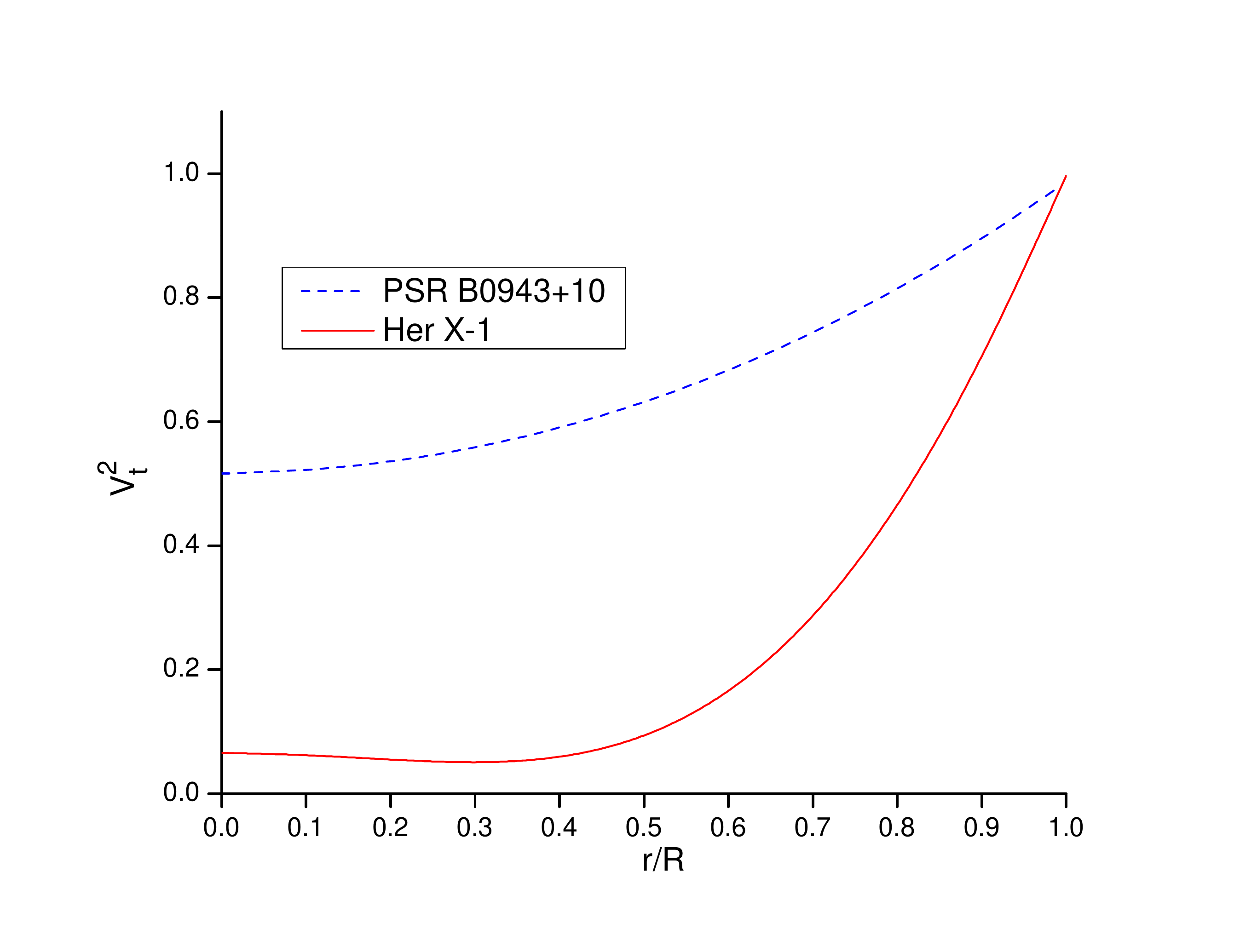}\\\includegraphics[width=5.5cm]{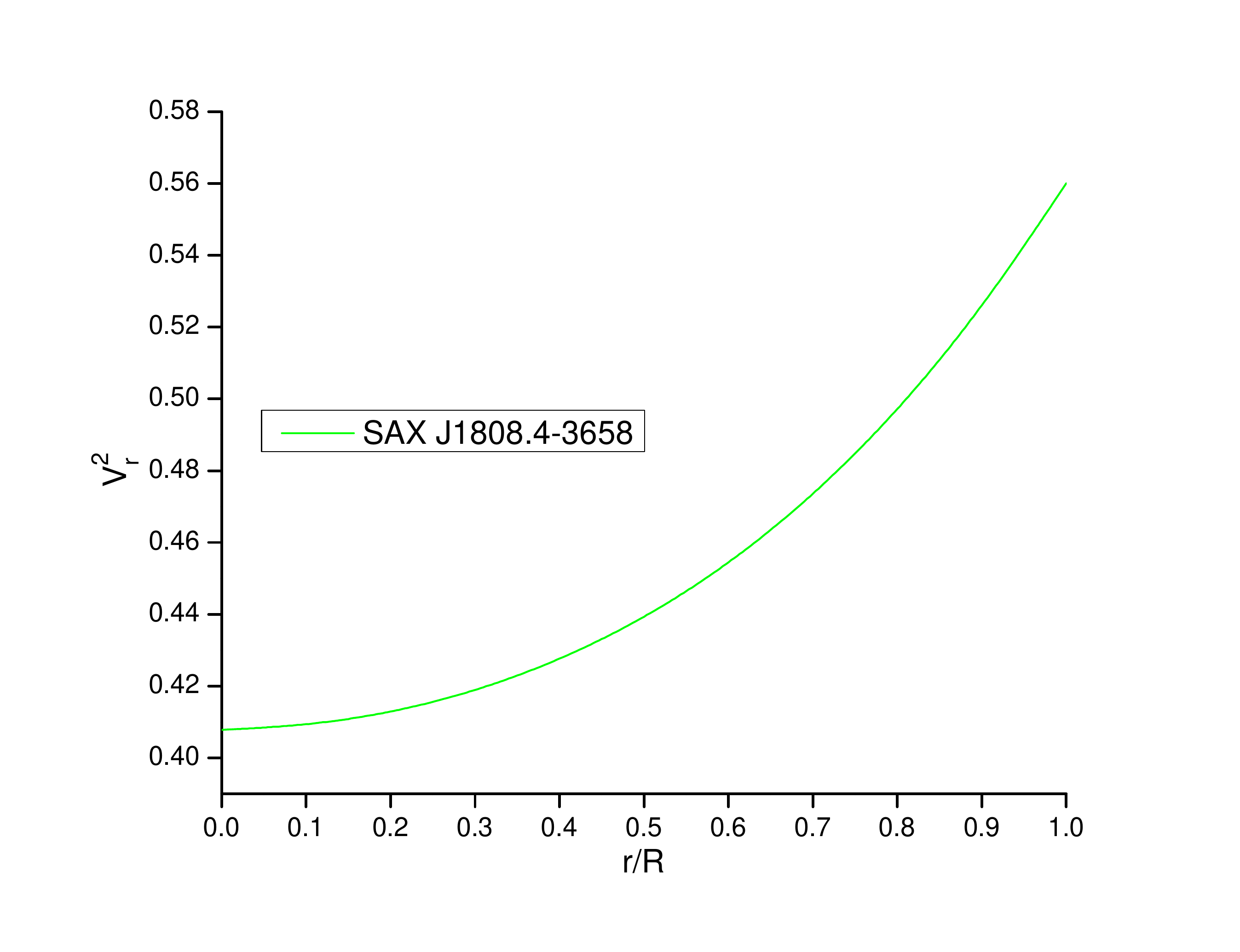}\includegraphics[width=5.5cm]{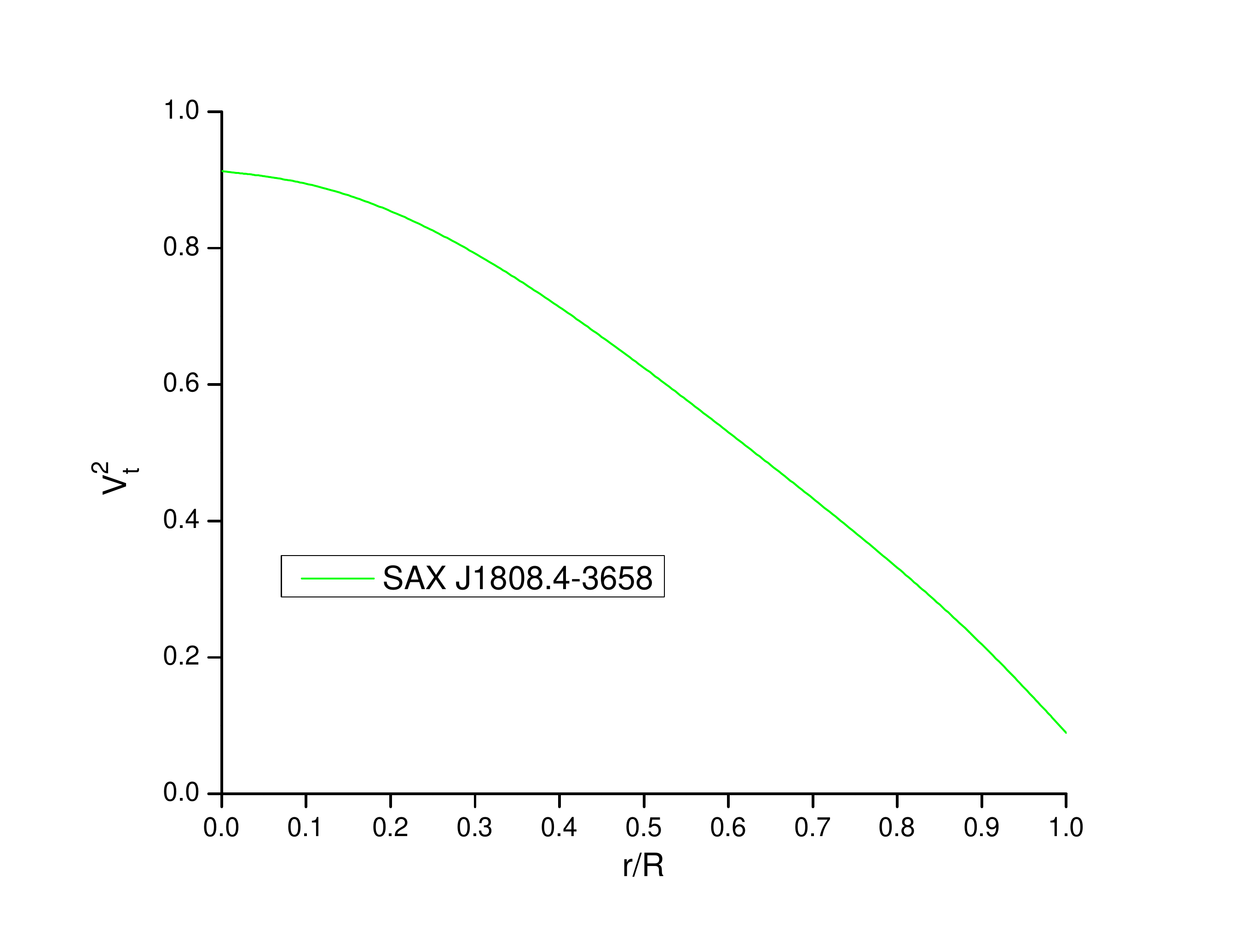}
\caption{Variation of radial and transverse velocity of sound with respect to radial coordinate $ r/R $ for $ K<0 $ (top) and $ K>1 $ (bottom). We have used same data set as Fig.\ref{f1}.}\label{f4}
\end{center}
\end{figure}

	The stability condition for relativistic stellar model is $ 0\leq V_{r}^{2},V_{t}^{2}\leq 1$. The upper bound of this inequality is enforce to avoid super-luminal extension which is called casuality condition and the lower bound break the dark energy fluctuation that grow exponential and carry out to non-physical state \cite{Pramit}. Here the expression of velocity of sound is as follows:
	\begin{eqnarray}
		V_{r}^{2} &=& \dfrac{dp_r}{d\rho} = \dfrac{\beta_1\,C^2(K-1)^2(3+Cr^2)^2 - 64\beta_2\,\pi^2\,K^2(1+Cr^2)^4}{C^2(K-1)^2(3+Cr^2)^2}\label{31}\\
		V_{t}^{2} &=& \dfrac{dp_t}{d\rho} \label{32}   
	\end{eqnarray}
	Here we have analyzed this condition through graphical representation due to the
	complexity of the expression of $ p_t $. In Fig.\ref{f4}, we plot the graph of radial and transverse velocity of sound for $ K<0 $(first row) and $ K>1 $(second row) of compact stars PSR B0943+10, Her X-1 and SAX J1808.4-3658.  Fig.\ref{f4} shows that the model satisfy the causality condition as well as Herrera cracking condition \cite{Herrera}. From Fig.\ref{f4}, we have observed that the velocity of sound is increasing for $ K<0 $. Also transverse velocity of sound is decreasing and radial velocity of sound is increasing for $ K>1 $. It is happened because of Buchdahl metric \cite{Maurya,Gupta,Gupta1}. Hence our solution is well behaved for discuss range of $ K $.
\begin{table}[h]
	\begin{center}
		\caption{The central pressure, central density, surface density, Buchdahl condition and  $ \beta_1 $  for compact star candidates.}\label{T2}
		\begin{tabular}{ccccccc}
			\hline
			Compact star& Central pressure & Central density & Surface density & Surface  & Buchdahl condition & $ \beta_1 $\\
			Candidates	& $ (dyne/cm^2) $ & $ (gm/cm^3) $ &  $(gm/cm^3) $ &Redshift $ (z_s) $ & $ (M/R\leq 4/9) $ & \\
			\hline
			PSR B0943+10 & $ 0.2911\times 10^{33} $ & $ 2.238\times 10^{13} $ & $ 2.187\times 10^{13} $ & 0.0116 & 0.014 & 0.32 \\
			\hline
			Her X-1& $ 4.332\times 10^{33} $ & $ 2.843\times 10^{13} $ & $ 1.873\times 10^{13} $ & 0.1507 & 0.1224 & 0.3\\
			\hline
			SAX J1808.4-3658& $ 2.741\times 10^{33} $ & $ 2.006\times 10^{13} $ & $ 1.356\times 10^{13} $ & 0.1166 & 0.0989 & 0.28 \\
			\hline
		\end{tabular}	
	\end{center}
\end{table}
\subsection{Energy conditions}
\begin{figure}[h]
\begin{center}
\includegraphics[width=5cm]{DEN}\includegraphics[width=5cm]{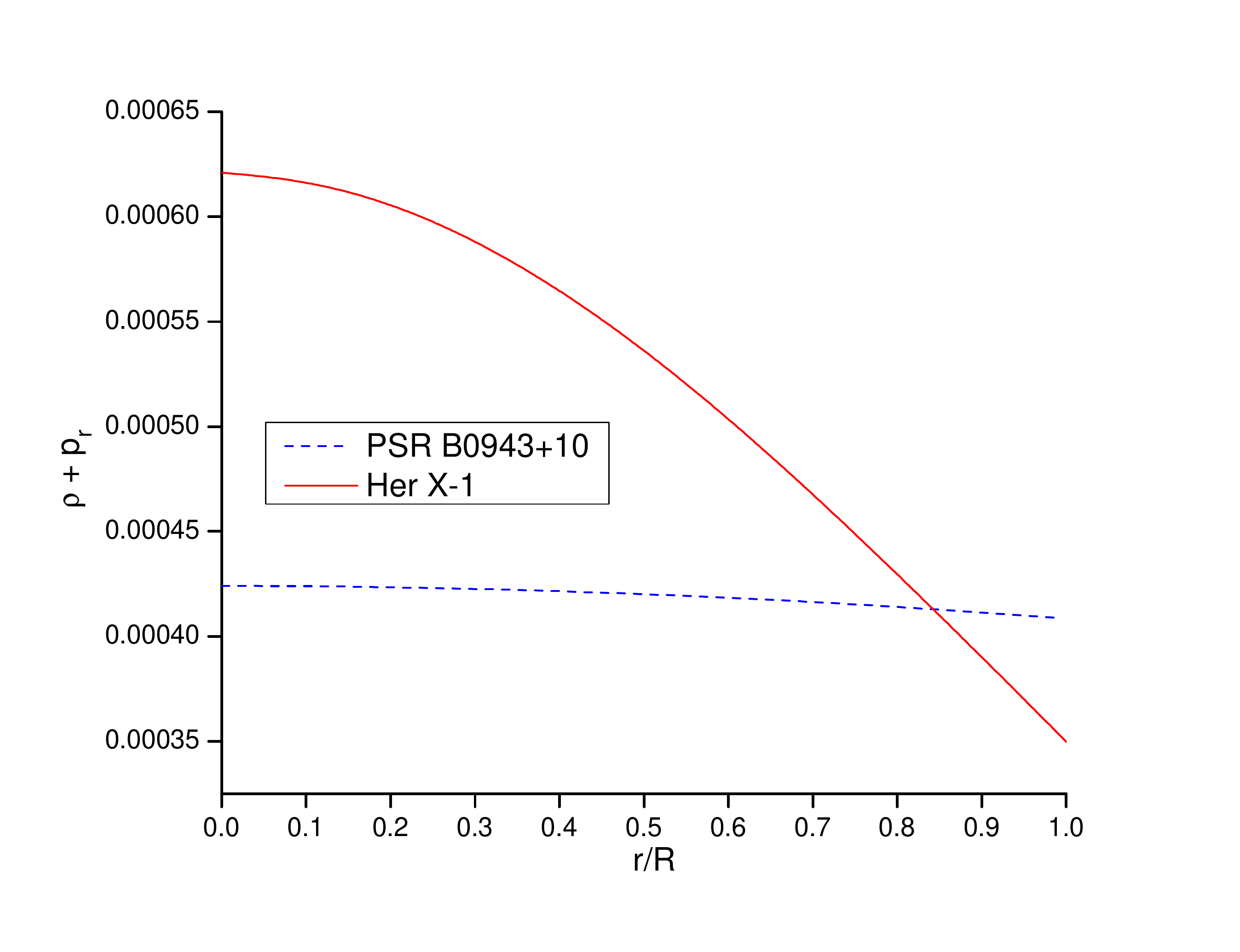}\includegraphics[width=5cm]{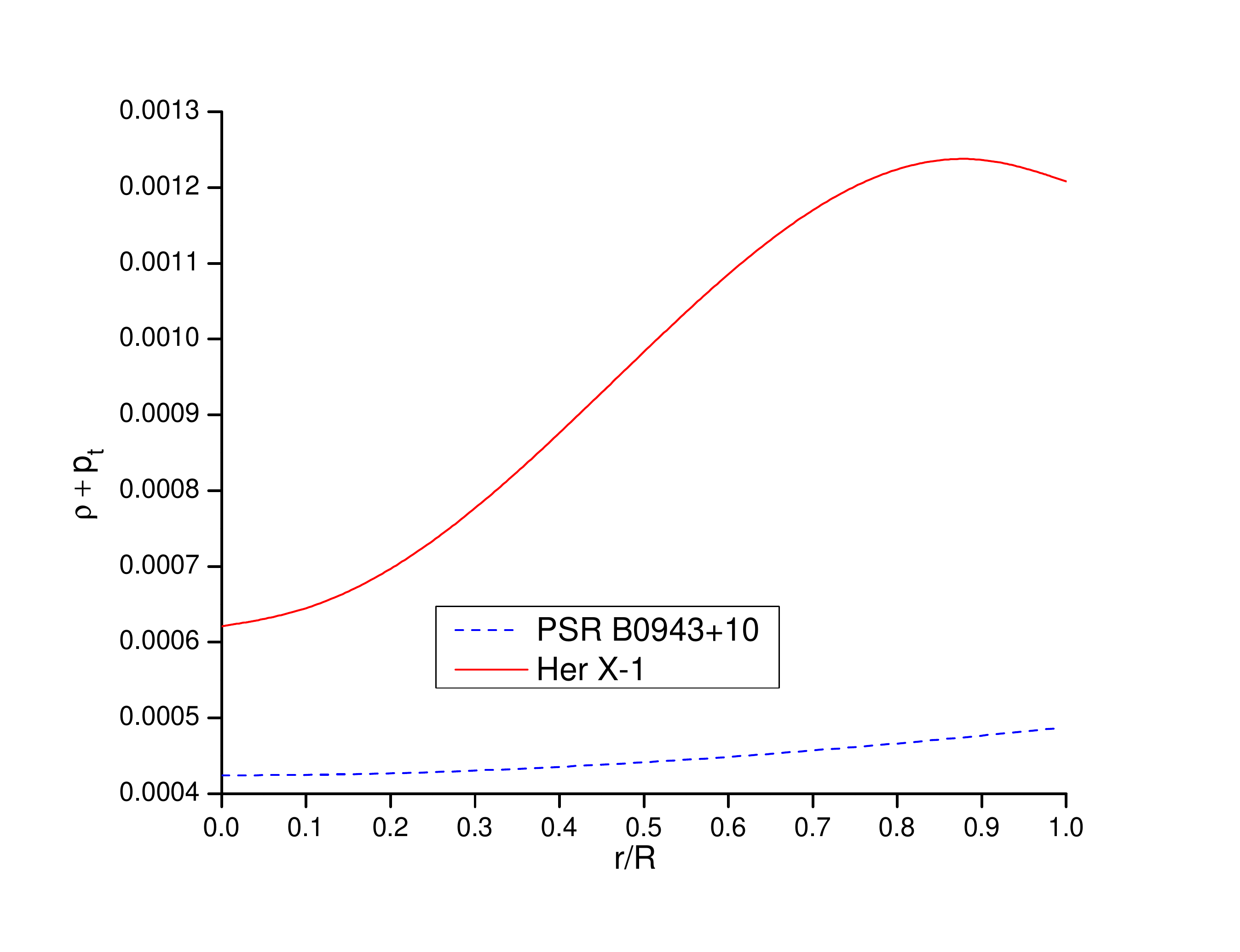}\\\includegraphics[width=5cm]{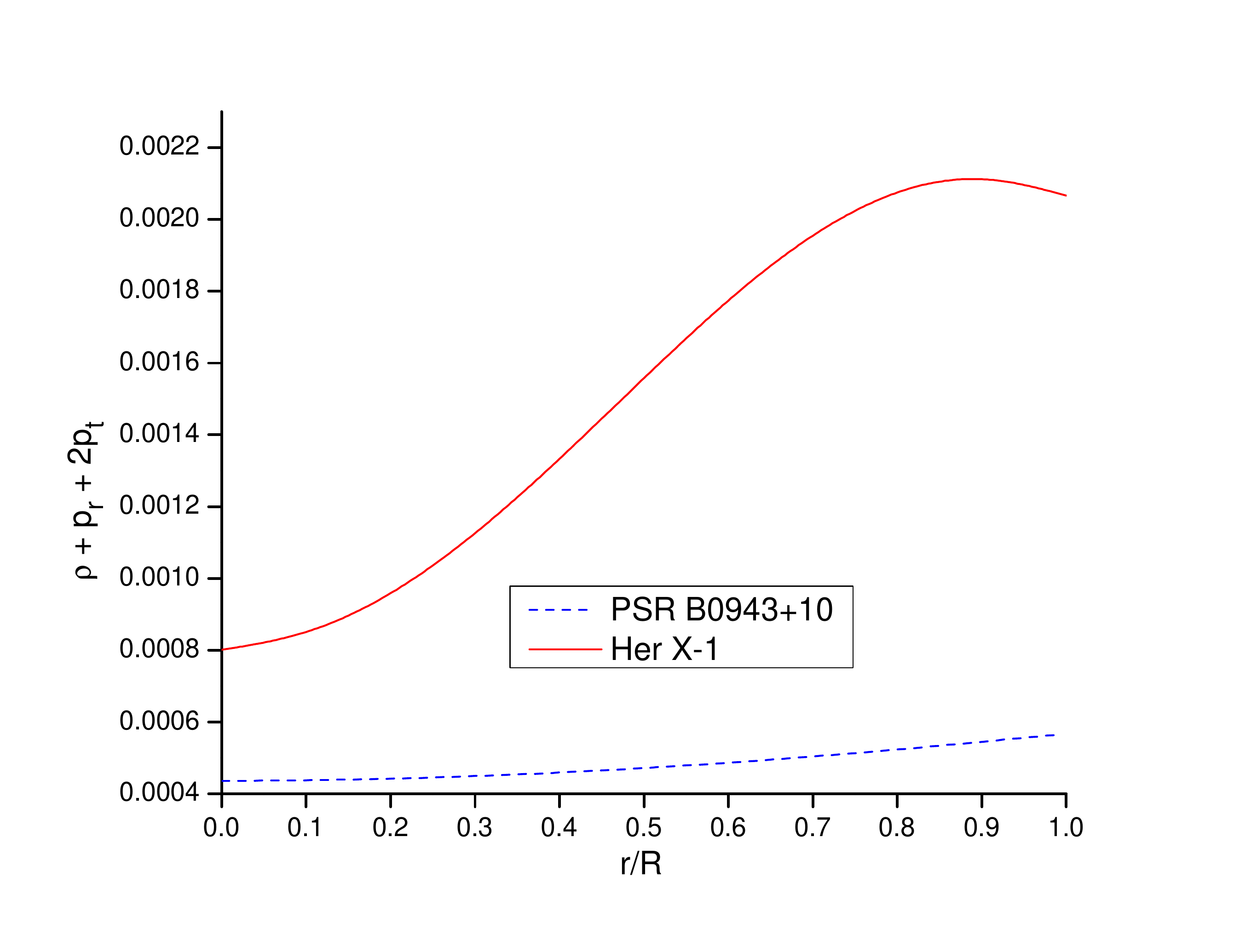}\includegraphics[width=5cm]{DEN1}\includegraphics[width=5cm]{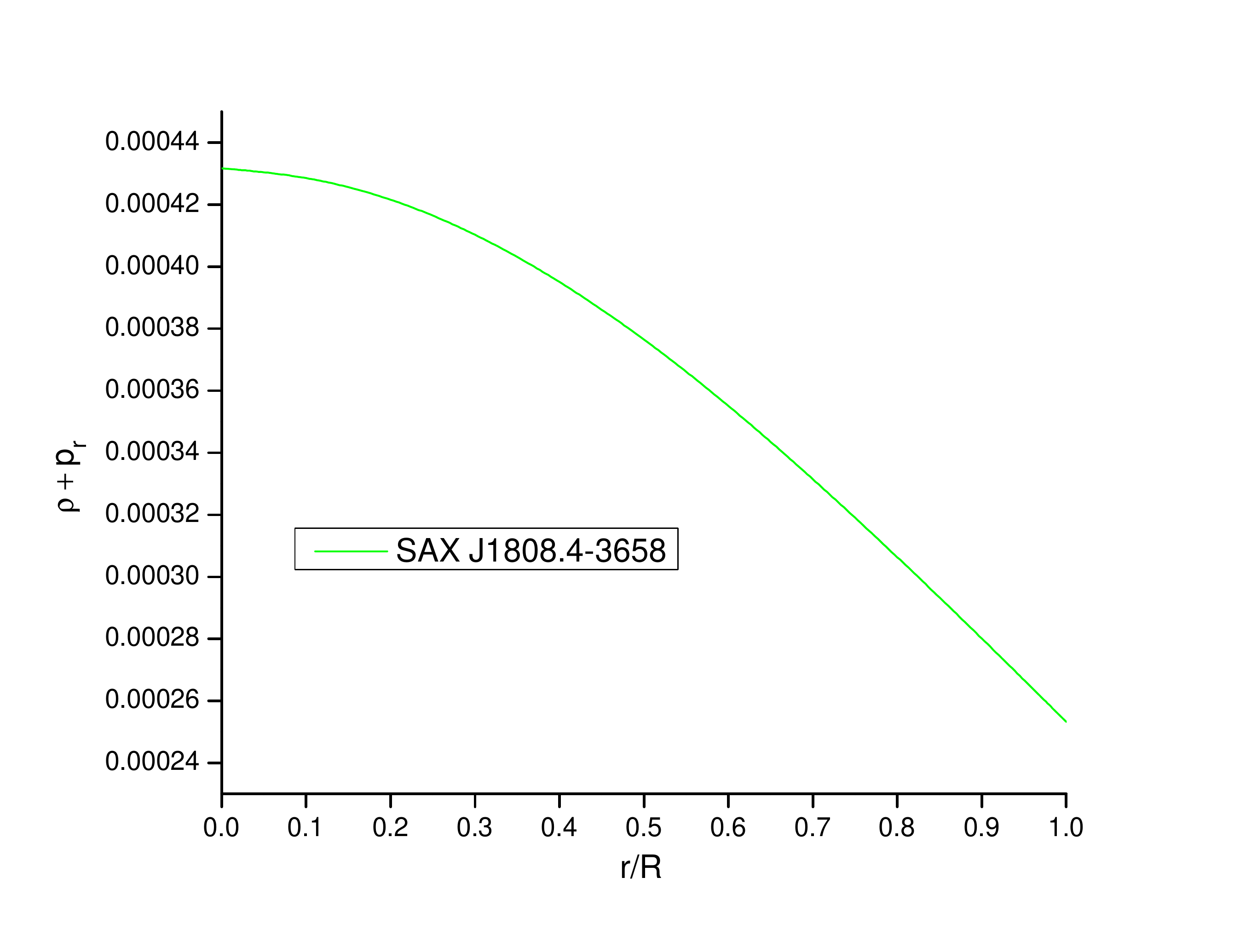}\\\includegraphics[width=5cm]{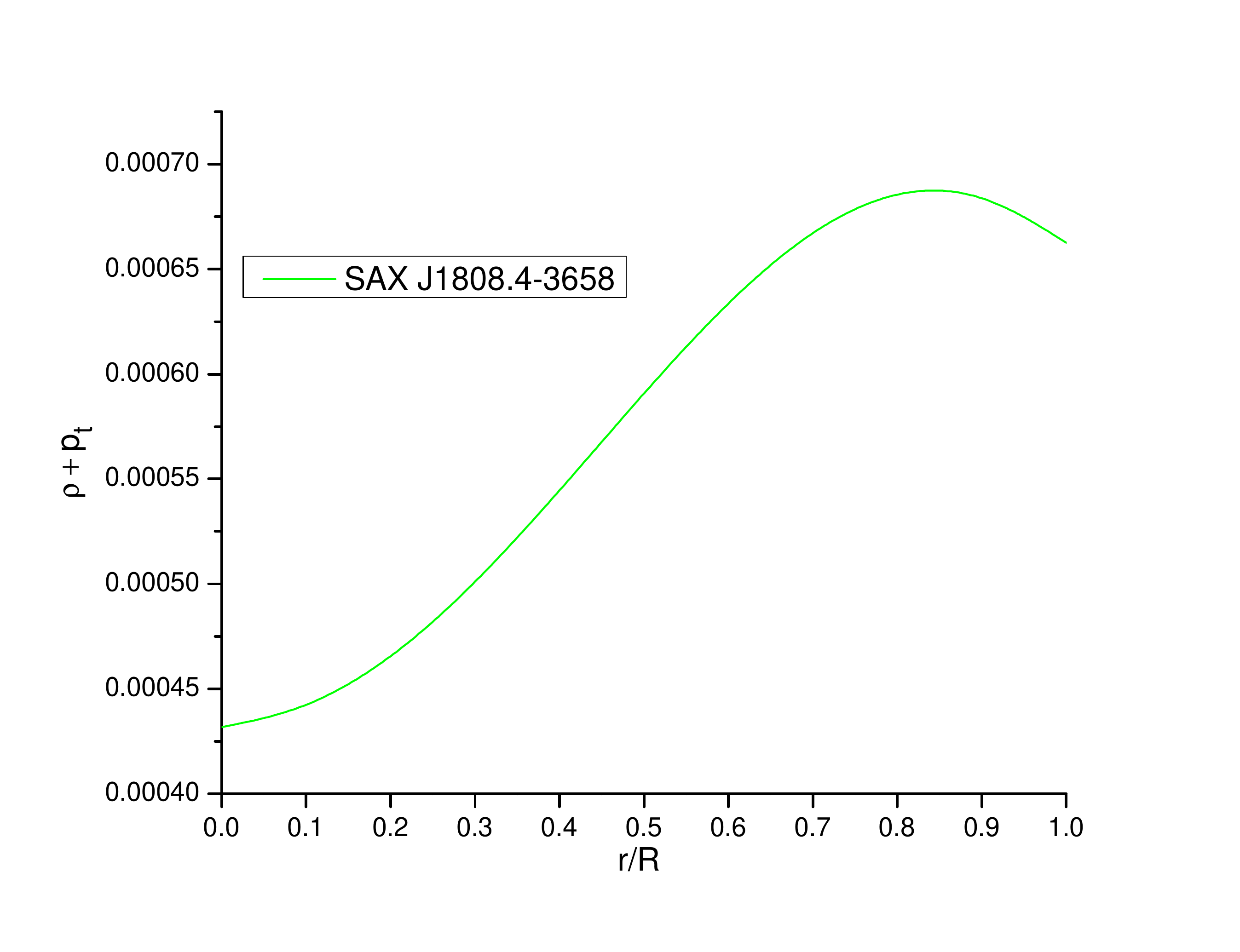}\includegraphics[width=5cm]{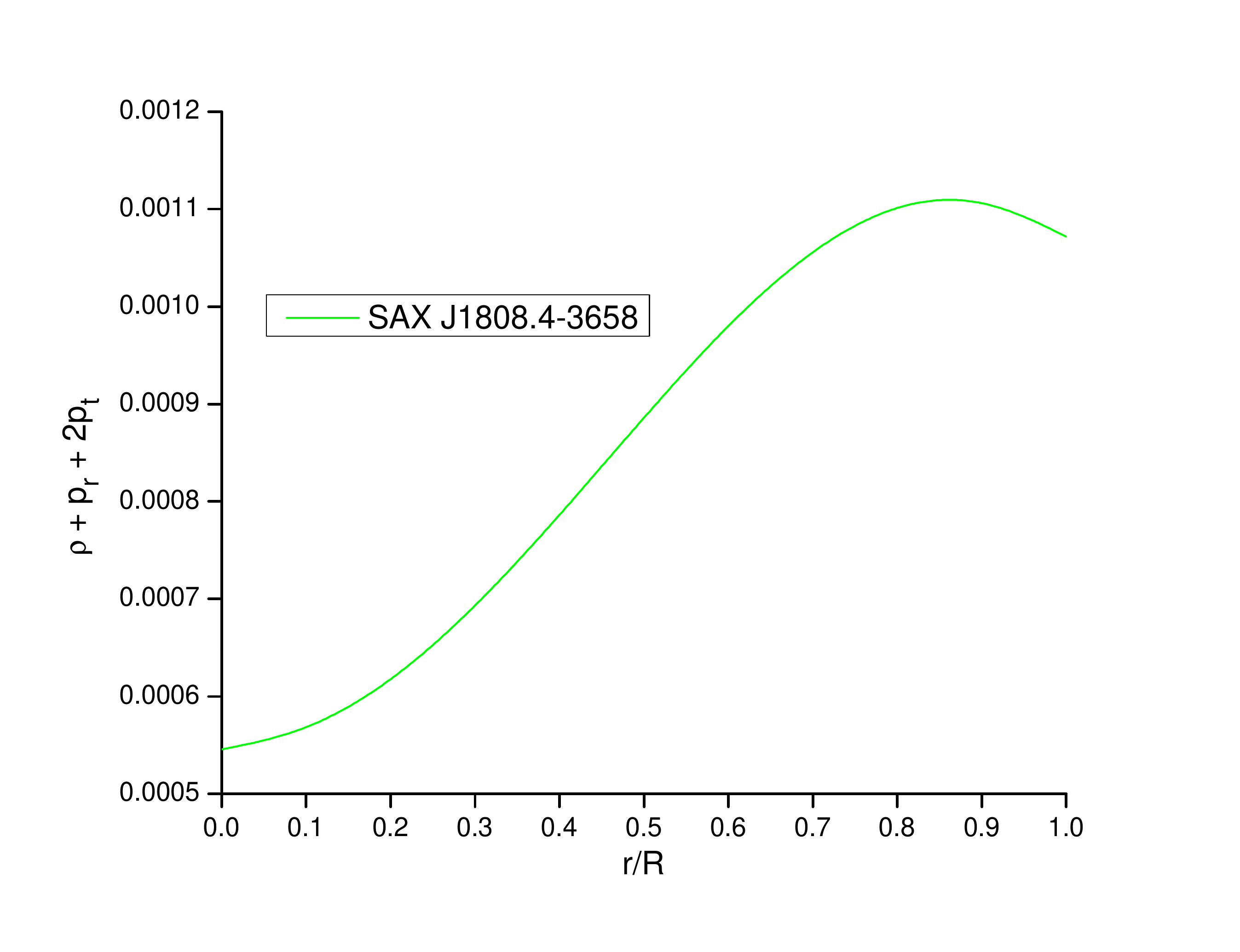}
\caption{The different energy conditions graphs for $ K<0 $ and $ K>1 $ have been plotted with respect to radial coordinate $ r/R $, where the first four graphs describe energy conditions corresponding to $ K<0 $ while the next four graphs correspond to $ K>1 $. For this, we have used same data set as Fig.\ref{f1}.}\label{f5}
\end{center}
\end{figure}

	In the context of general relativity, the relativistic anisotropic compact star models will be physically acceptable if it satisfy the energy conditions i.e, null energy condition (NEC), weak energy condition (WEC) and strong energy condition (SEC). It also plays an important role to understand the nature of matter distribution \cite{Pramit,Gasperini}. These condition are defined as \cite{Ponce,Visser}
	\begin{enumerate}[(i)]
		\item (NEC) $\rho \geq 0 $
		\item (WEC) $\rho + p_r\geq 0 $,\,\,\, $\rho + p_t\geq 0 $
		\item (SEC) $\rho + p_r + 2p_t\geq 0 $
	\end{enumerate} 
	If the anisotropic compact stars satisfy the above inequalities then one can say that the energy momentum tensor is positive within the configuration. According to Maurya \cite{Maurya1}, violation of energy conditions implies the unphysical stress energy tensors. Here, our anisotropic model satisfies all the above energy conditions, it is shown in Fig.\ref{f5}. So our model has a well-behaved and positive energy momentum tensor.
	\subsection{Adiabatic index}
\begin{figure}[h]
		\begin{center}
			\includegraphics[width=5cm]{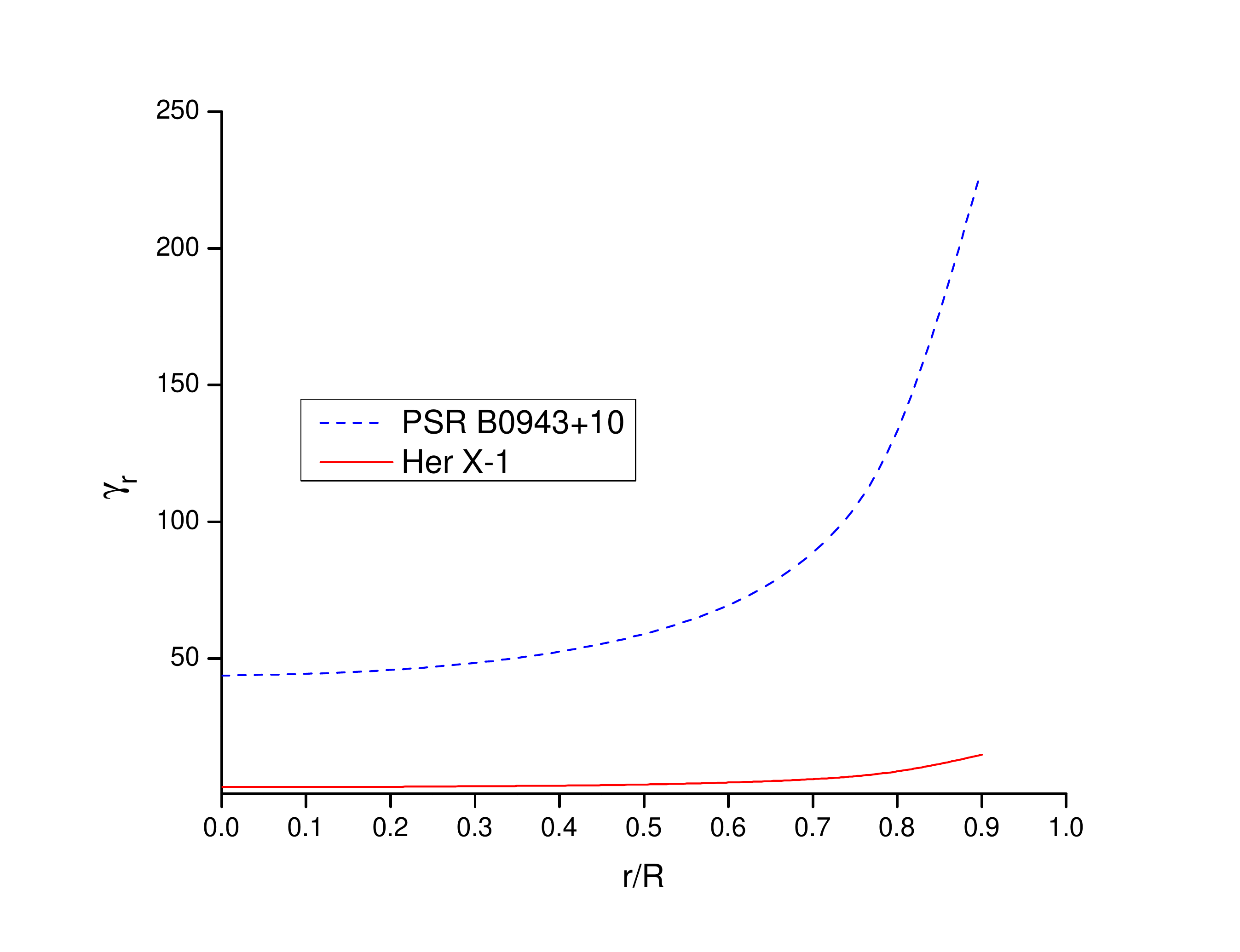}\includegraphics[width=5cm]{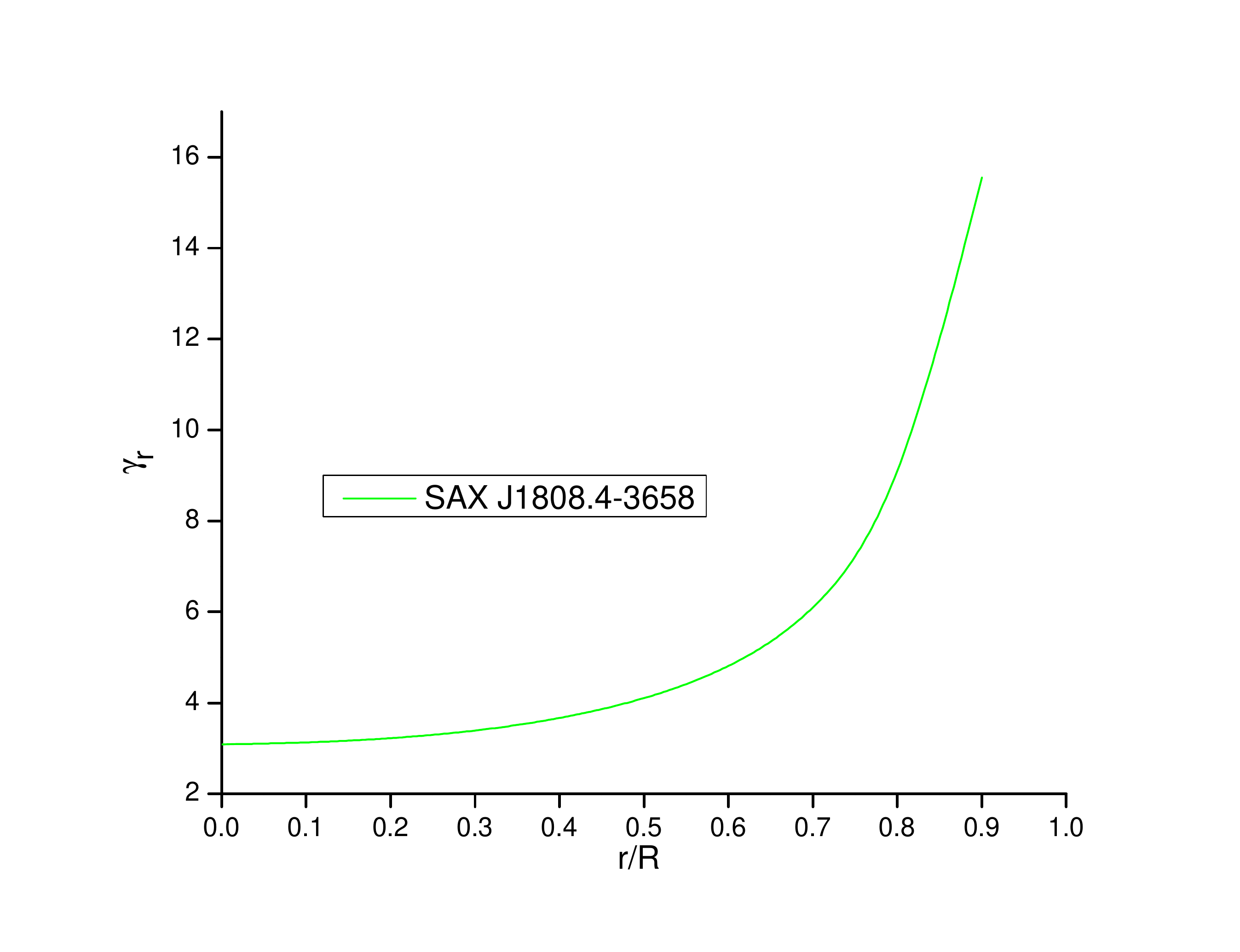}
			\caption{Variation of Adiabatic index ($ \gamma_r $) versus radial coordinate $ r/R $ for $ K<0 $(left) and $ K>1 $(right). For plotting these graphs, We have used same data set as Fig.\ref{f1}. }\label{f6}
		\end{center}
	\end{figure}

	The stability of the relativistic as well as non relativistic compact object
	likewise relies on the two specific heats given by $ \gamma $. Heintzmann and
	Hillebrandt \cite{Heintzmann} recommended that the adiabatic index $ \gamma $ must be more than $ \frac{4}{3} $ at all interior point of relativistic anisotropic compact object. In other side  Bondi\cite{Bondi} point out that the model is unstable  for $ \gamma <\frac{4}{3}$. For an anisotropic relativistic compact stars, the adiabatic index is given by  
	\begin{equation}
		\gamma_r = \Big(\dfrac{\rho + p_r}{p_r}\Big)\dfrac{dp_r}{d\rho}\label{33}
	\end{equation}
	We have drawn the graph of radial adiabatic index in Fig.\ref{f6}. The figure shows that the adiabatic index greater than $ \frac{4}{3} $  for each compact stars. Hence we can say that our model is stable.
	\subsection{ Harrison–Zeldovich–Novikov stability criterion}
\begin{figure}[h]
		\begin{center}
		\includegraphics[width=5.5cm]{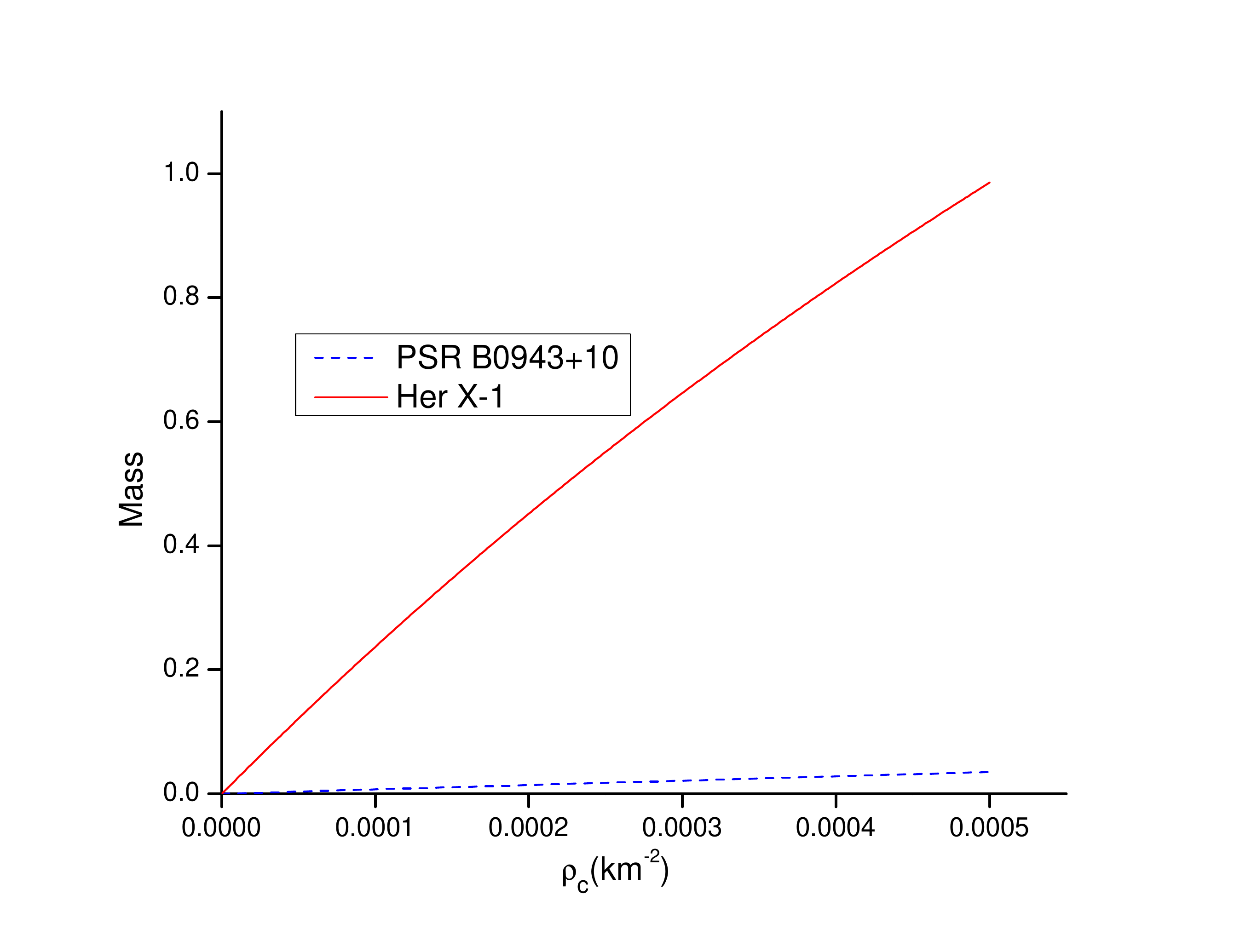}\includegraphics[width=5.5cm]{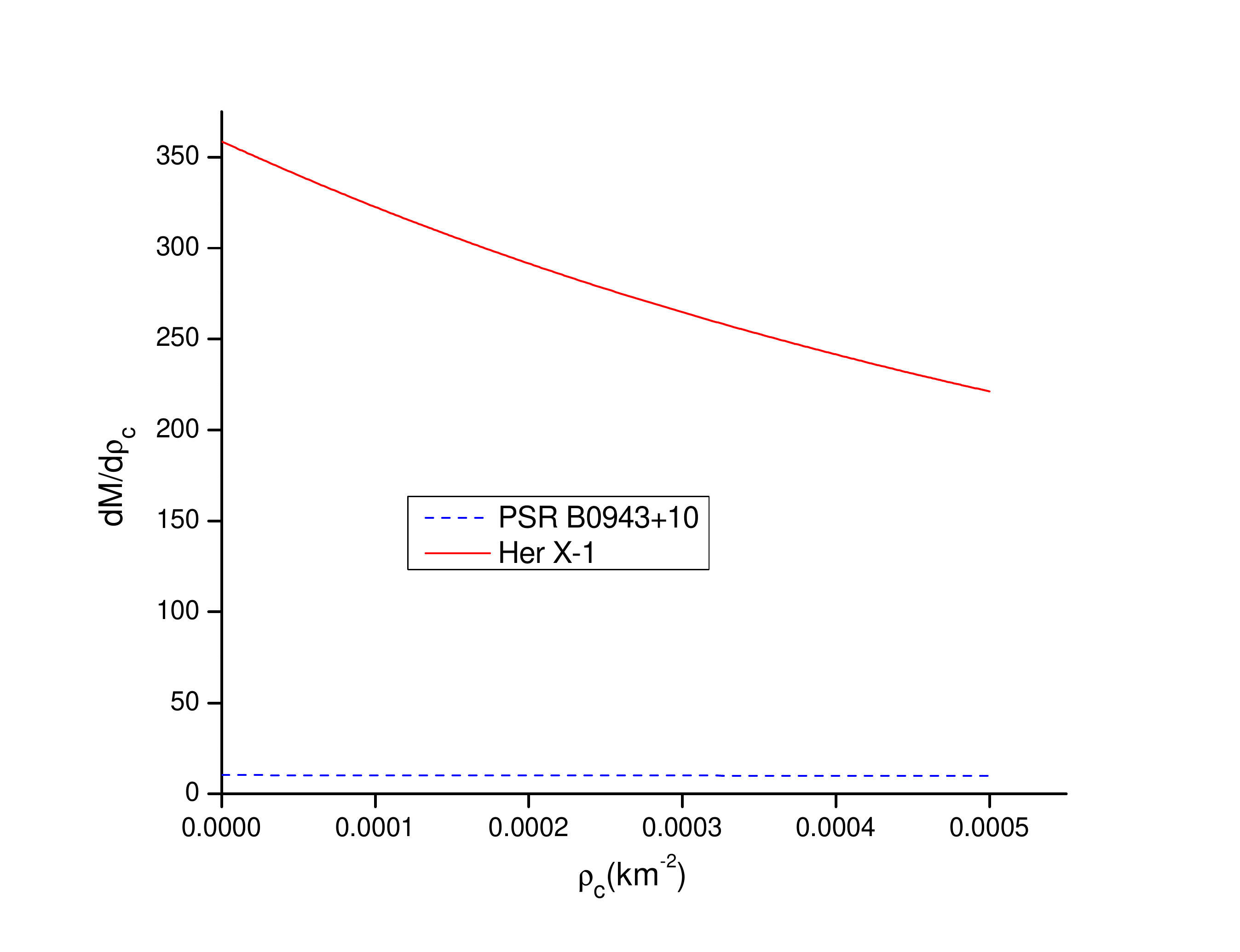}\\
		\includegraphics[width=5.5cm]{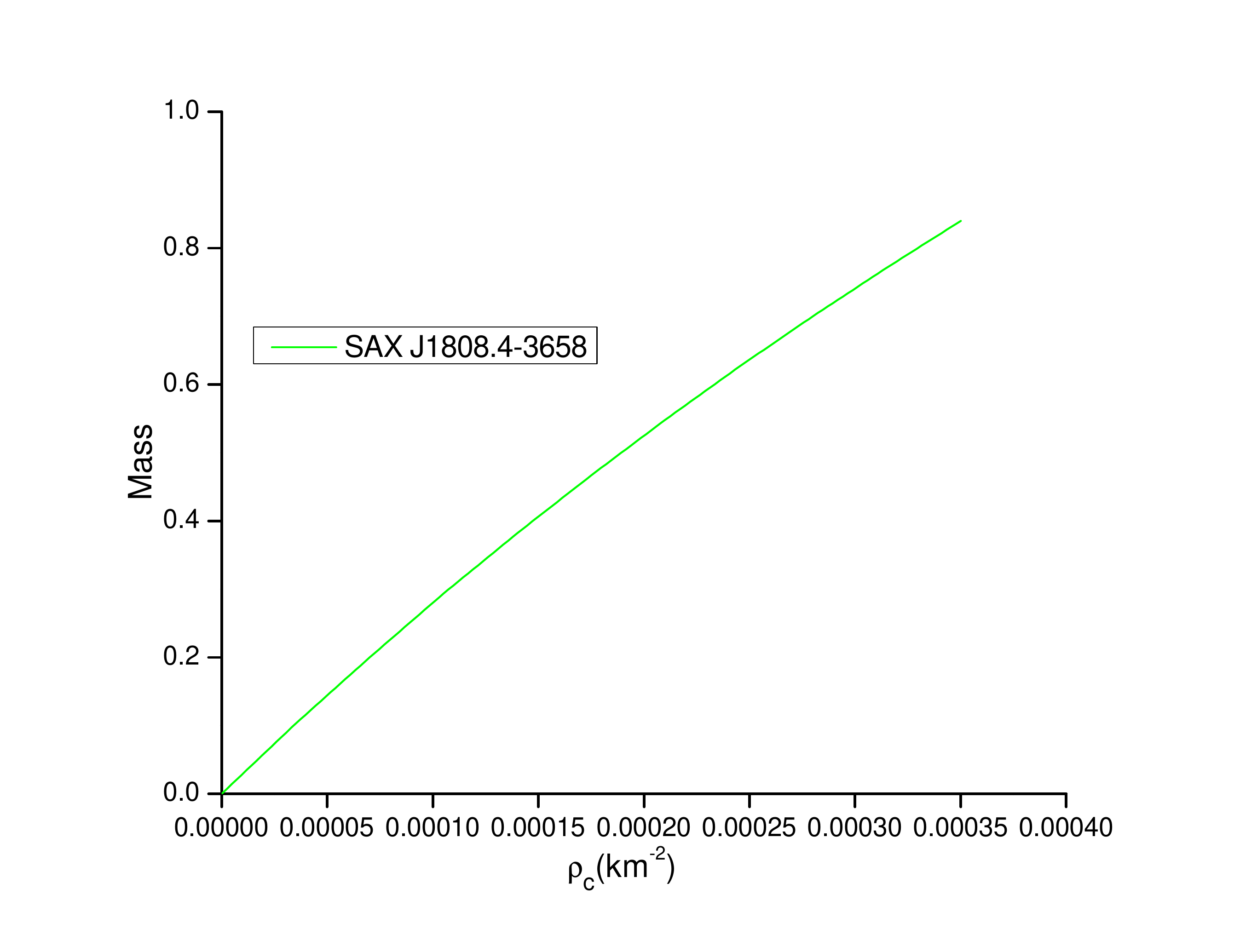}\includegraphics[width=5.5cm]{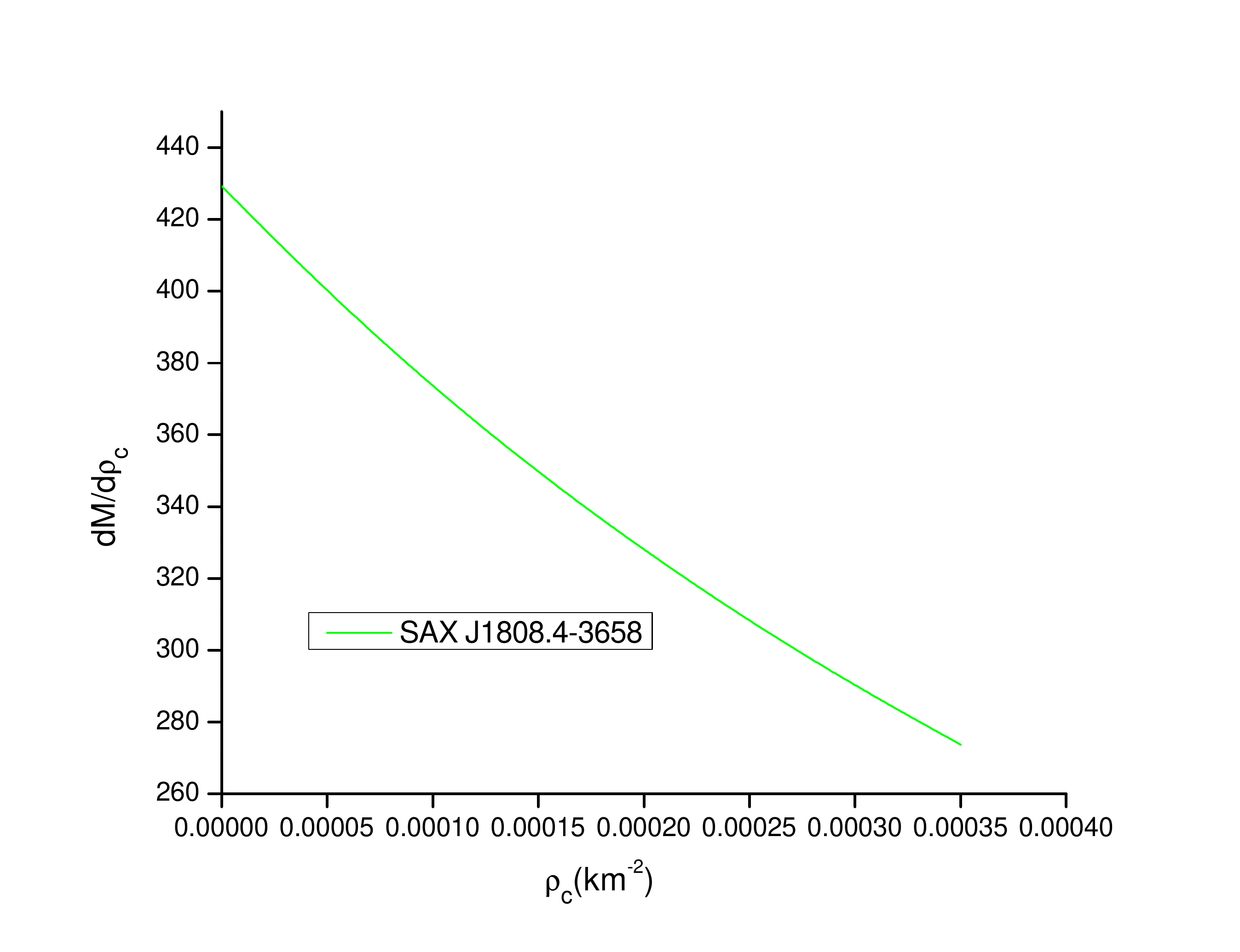}
		\caption{Variation of mass and  ($ dM/d\rho_c $) versus central density $ \rho_c $ for $ K<0 $(left) and $ K>1 $(right).}\label{f7}
		\end{center}
	\end{figure}

	The stability criterion demonstrate that the solution is static and stable under an infinitesimal radial perturbation.  In this criterion, it is postulate that the any stellar configuration has an increasing mass with increasing central density, i.e. $ dM/d\rho_{c} > 0 $ represents stable configuration and vice versa \cite{Harrison,Zeldovich}. If the mass remains constant with increasing central density, i.e. $ dM/d\rho_{c} = 0 $, we get the turning point between stable and unstable region. For our model, we obtained $ M(R) $ and $ dM/d\rho_{c} $ as follows-\begin{equation}
		M(\rho_c)=\frac{4\,\pi\,\rho_c\, R^3\,(K-1)}{(3(K-1)+8\,\pi\,K\,\rho_c\,R^2)}~~~~~\textrm{and}~~~~\frac{dM}{d\rho_c}=\frac{12\,\pi\,R^3\,(K-1)^2}{(3(K-1)+8\,\pi\,K\,\rho_c\,R^2)^2}\label{34}
	\end{equation}
	Hence from Fig.\ref{f7}, we can see that the mass is increase with the increment of the central density. Hence we conclude that the model represents static and stable configuration.
	\subsection{Mass function and Compactness}
\begin{figure}[H]
		\begin{center}	\includegraphics[width=5.5cm]{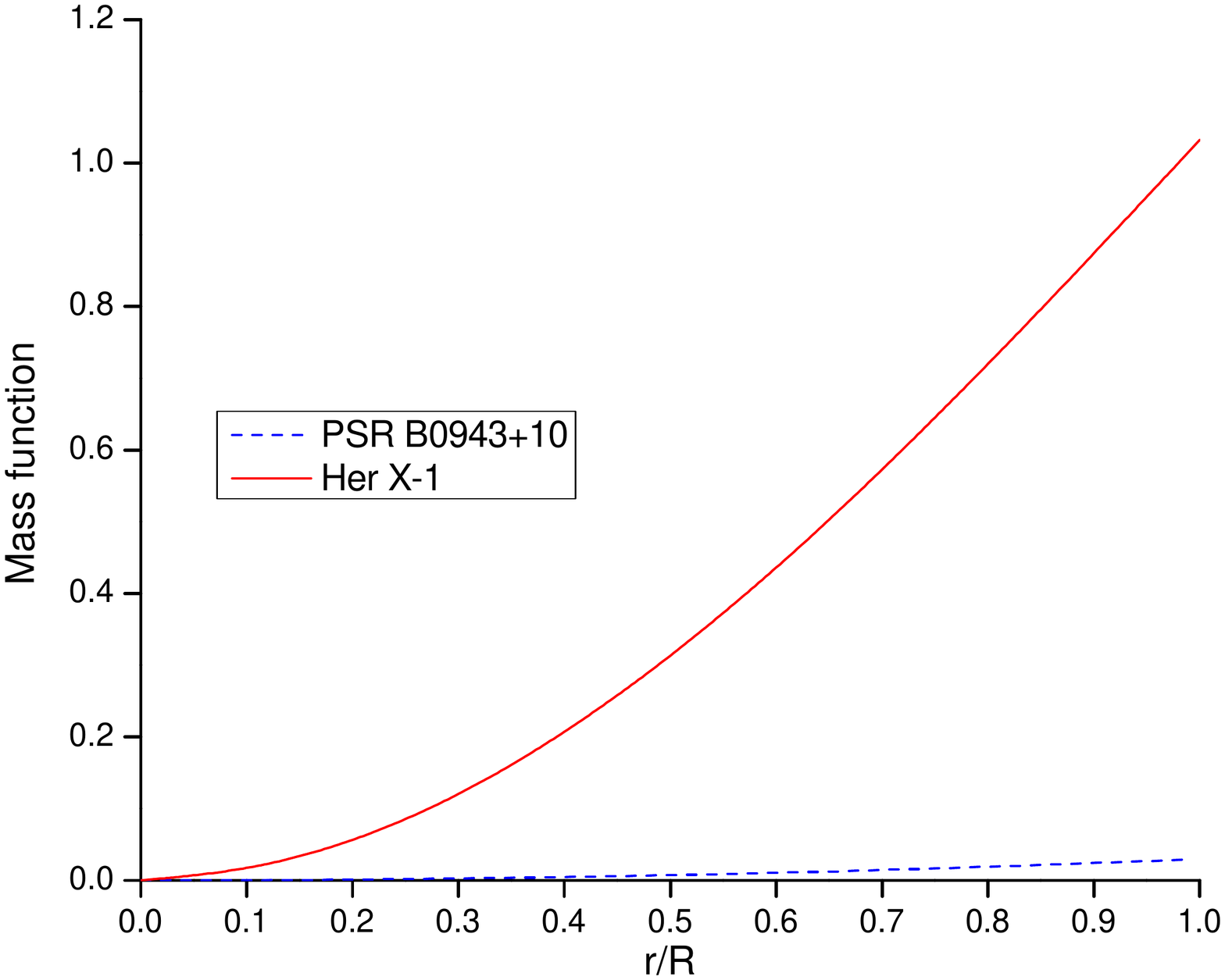}\includegraphics[width=5.5cm]{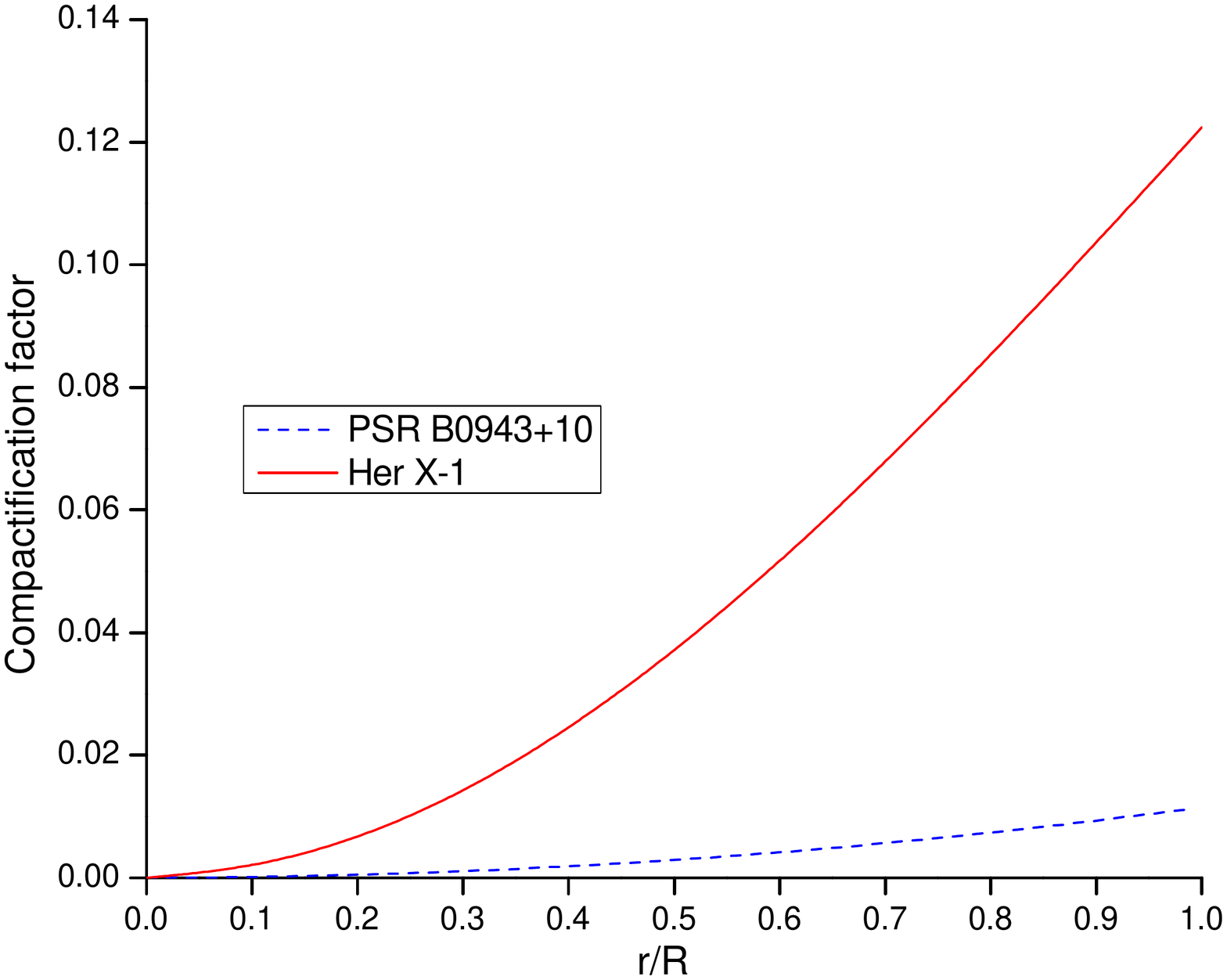}\\\includegraphics[width=5.5cm]{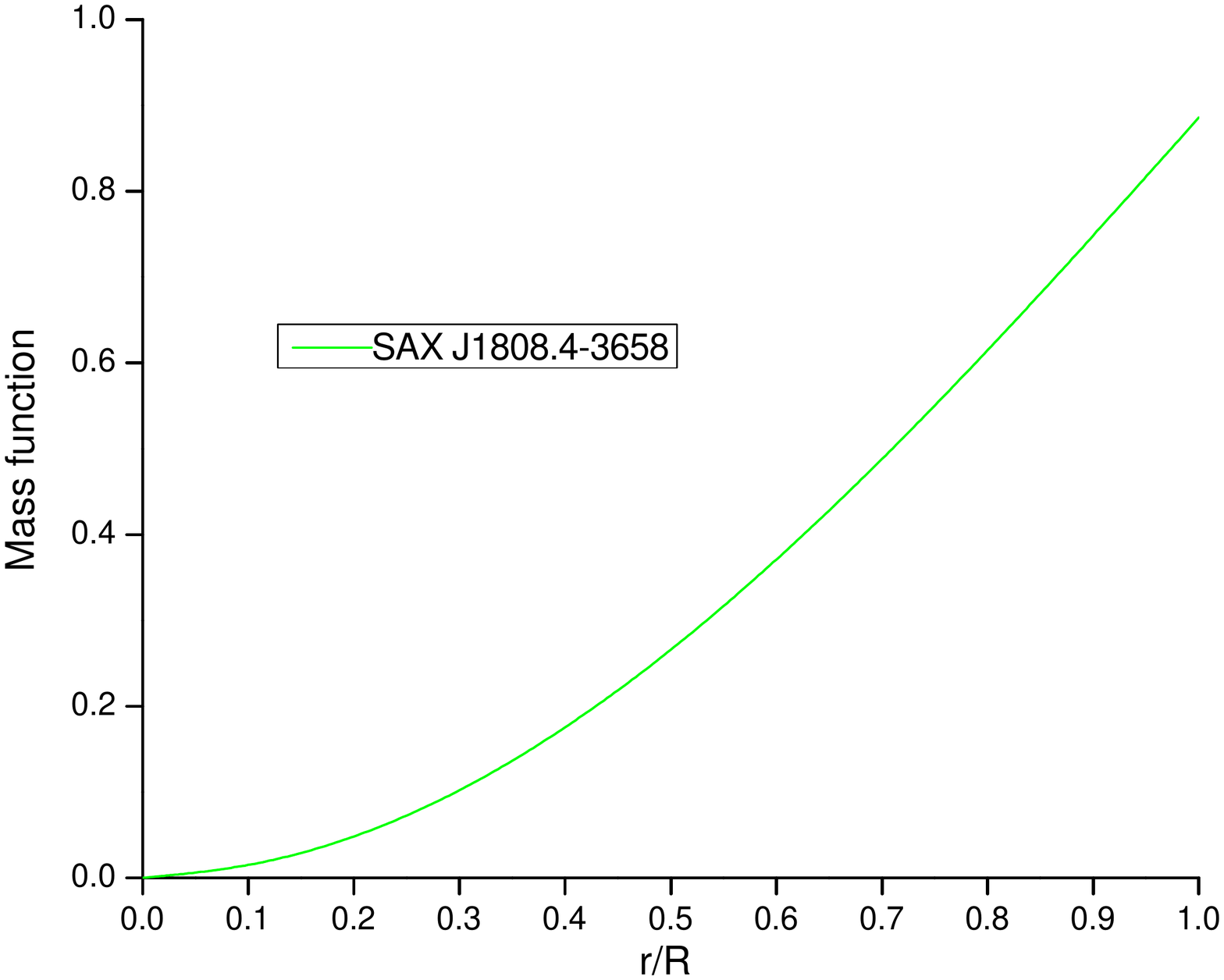}\includegraphics[width=5.5cm]{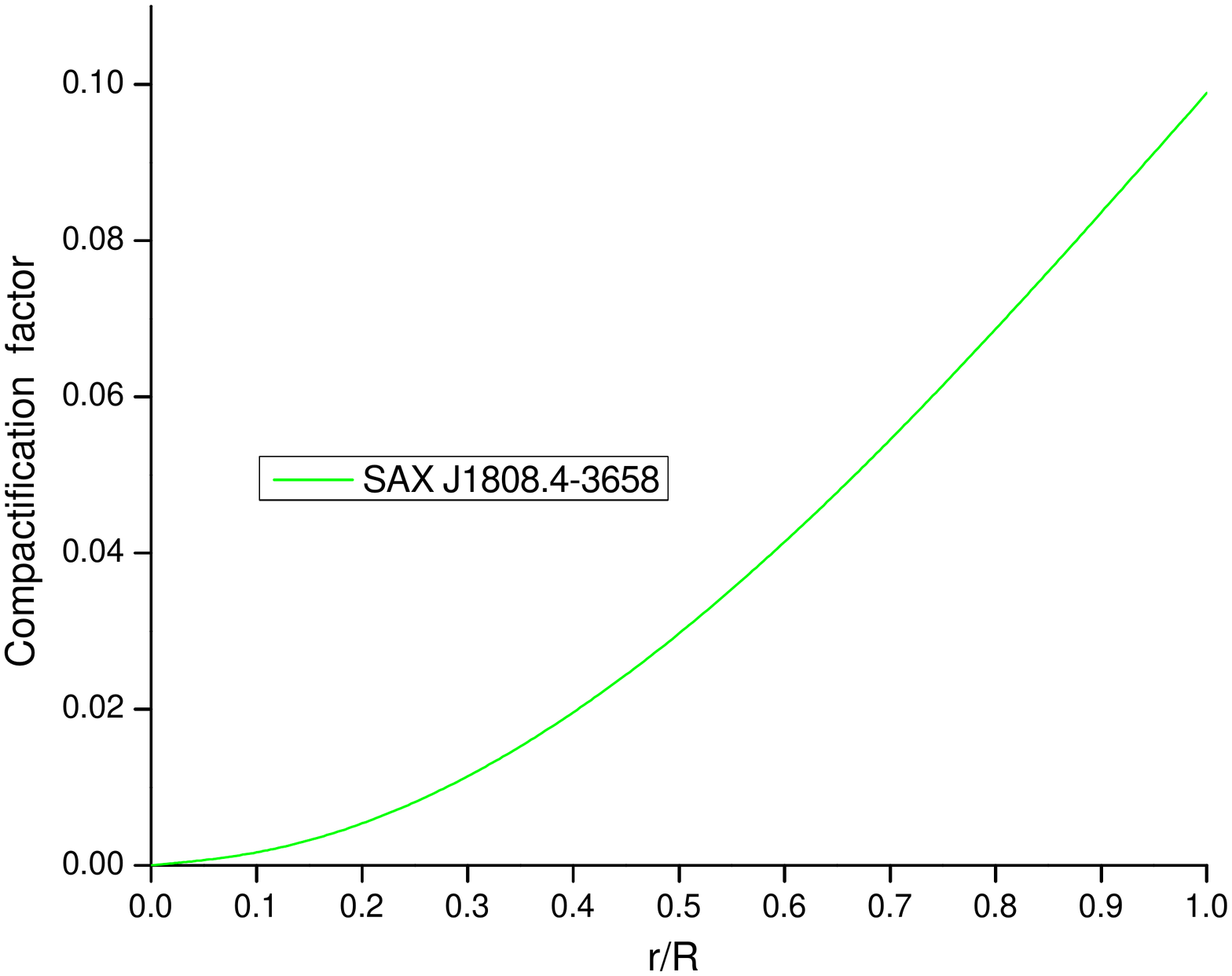}
			\caption{Variation of the mass function and compactification factor with respect to the radial coordinate $ r/R $ for $ K<0 $ (top) and $ K>1 $ (bottom). For plotting these graphs, we have used same data set as Fig.\ref{f1}.}\label{f8}
		\end{center}
	\end{figure}

	In this subsection, we have discussed the mass function and  mass-radius relationship. In this context, Buchdahl [19] suggested that the mass-radius ratio of a relativistic static spherically symmetric fluid stellar model should be $ \frac{M}{R} \leq \frac{4}{9} $. In this regards, Mak and Harko \cite{Mak} have obtained a generalized formula for the mass-radius ratio. In our model, we have obtain the relationship of the mass function of relativistic compact stars as follows 
	\begin{eqnarray} 
		M(r)= \dfrac{Cr^3(K-1)}{2K\,(1+Cr^2)},\label{35}
	\end{eqnarray}
	In Fig.\ref{f8}, we can see that the mass function is regular at the center of the stars.  Also it is monotonic increasing function of $ r $ and positive inside the relativistic compact stars. \par
	The ratio of the mass to the radius of a strange star known as the compactness factor. The expression of compactness factor of this model given as
	\begin{eqnarray}
		u = \dfrac{M}{r} =	\dfrac{Cr^2(K-1)}{2K\,(1+Cr^2)}\label{36}
	\end{eqnarray}
	The profile of $ u(r) $ is shown in Fig.\ref{f8}. From this figure, we can see that compactness factor $ u(r) $  increases with increase $ r/R $. This shows that the compactness of compact stars lies in the expected range of Buchdahl limit\cite{Buchdahl}.\par
	We have determined the surface redshift from compactness $ u $, which is given by 
	\begin{eqnarray}
		z_s = (e^{\lambda(R)/2}) -1 = (1-2u)^{-1/2} -1\label{37}
	\end{eqnarray}
	From equation (\ref{37}), we get 
	\begin{eqnarray}
		z_s = \Bigg(\dfrac{K+CR^2}{K(1+CR^2)} \Bigg)^{-1/2} -1\label{38}
	\end{eqnarray}
	The numerical values of surface redshift for different compact stars are given in Table-\ref{T2}. According to Straumann \cite{Strauman} and Buchdahl \cite{Buchdahl}, the value of surface redshift in absence of a cosmological constant for isotropic stellar is given by $ z_s\leq 2. $ However, Karmakar and Barraco \cite{Karmakar,Barraco} have suggested that the surface redshift for anisotropic stellar model could be 3.84. Also, Boehmer and Harko \cite{Boehmer} showed that the value of surface redshift increased up to $ z_s \leq 5. $ The maximum value of surface redshift of our model is ($ 0.1507$).
	
	\subsection{Moment of Inertia}
\begin{figure}[H]
\begin{center}	\includegraphics[width=5.5cm]{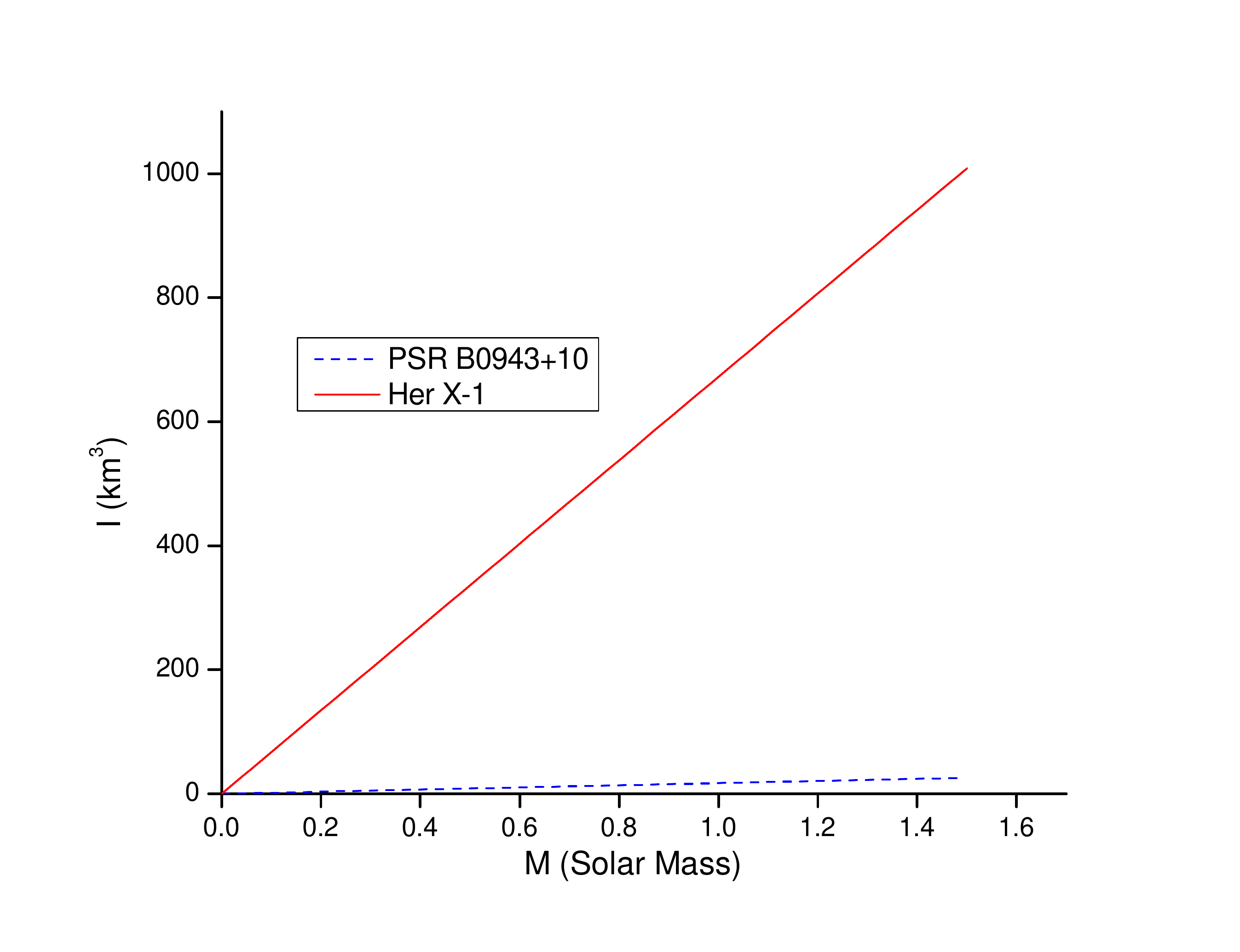}\includegraphics[width=5.5cm]{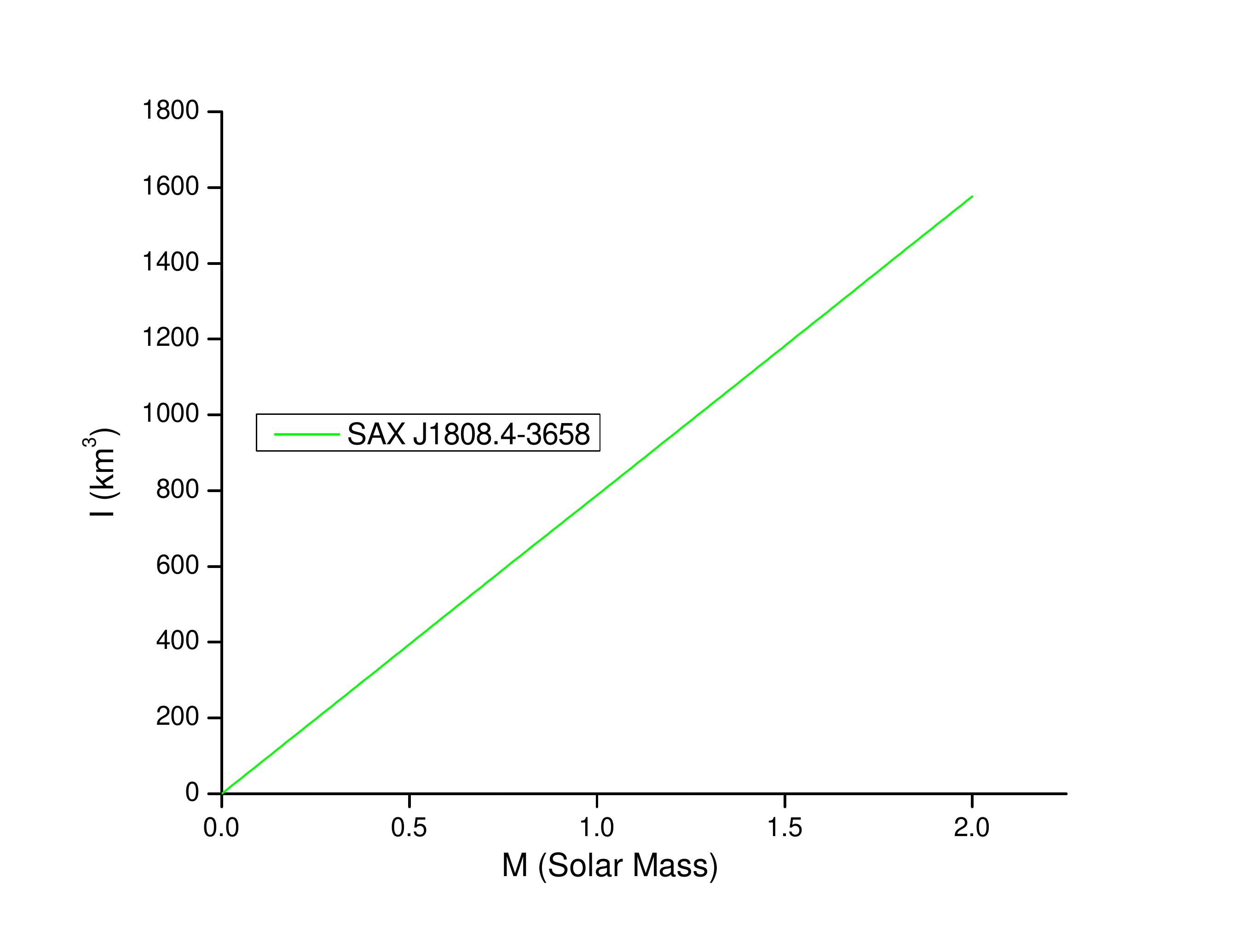}
\caption{Variation of moment of inertia $ I $ with respect to the mass $ M_{\odot} $ for $ K<0 $ (left) and $ K>1 $ (right). For plotting these graphs, we have used same data set as Fig.\ref{f1}.}\label{f9}
\end{center}
\end{figure}

	The moment of inertia of compact stars is the most sensitive to the dense matter equation of state, which could be utilized to constrain theoretical models \cite{Bejger}. In present model, we consider Bejger-Haensel method \cite{Maurya2,Bejger} based approximate formula for computing the moment of inertia which is given as
	\begin{eqnarray}
		I = \dfrac{2}{5}[1+\xi]MR^2, \label{39} 
	\end{eqnarray}
	where $ \xi = (M/M_{\odot})(km/R) $. Using above equation we analyzed the behavior of moment of inertia which is represented in Fig.\ref{f9}. From Fig.(\ref{f9}), we can conclude that the sensitivity of $ I - M $ graph increase and rigidity of the state equation is better for all compact stars listed in Tabel-\ref{T1}.
\section{Conclusion} \label{Sec6}
In this work we have obtained a new class of well-behaved anisotropic compact star model after prescribing a suitable metric potential and equation of state. In our model the interior spacetime meets the exterior Schwarzschild spacetime smoothly. The pressure, density and the ratio $ \dfrac{p}{\rho }$ are, as per expectation, seen to be monotonically decreasing towards the surface for $K=-1.52,\,\,-11.52$ and $ K=14.50 $. The anisotropic factor $\Delta$ is graphically represented in Fig. \ref{f2} and found to be positive implying that the anisotropic force being repulsive in nature is beneficial for the stability of our model. Also, our model satisfies a vital criteria of having $\Delta = 0$ at the center of the star for a model attempting to represent realistic stellar objects. We have taken generalized TOV equation to demonstrate hydrostatic stability of our model. The energy conditions viz. null energy conditions, weak energy conditions and strong energy conditions are satisfied by our model (see Fig.\ref{f5}). The redshift is also decreasing from the center to surface for the values of $ K $  mentioned as above. For analyzing the stability of our model we have shown that the adiabatic index is greater that $\frac{4}{3}$ and represented graphically that our model obeys the Harrison–Zeldovitch–Novikov criterion (See Fig. \ref{f7}). We calculated the mass functions and moment of inertia for our candidate stars to underscore the compactness of the model and choice of our EoS. The model also satisfies reality and causality conditions i.e, $(dp/c^{2}d\rho)<1 $ everywhere inside a compact star. We have verified that the stars like PSR B0943+10 for $ K=-1.52 $, Her X-1 for $ K=-11.52 $ and SAX J1808.4-3658 for $ K=14.50 $ to be close candidates for our proposed model. The behaviors of various physical parameters are demonstrated in Figs.\ref{f1}-\ref{f9}. The numerical values of physical quantities are shown in the Tables \ref{T1}-\ref{T2} where the various symbols used in the table are as follows:
$ z_s= $ redshift at the surface, solar mass $ M_{\odot}=1.475km ,\,\, G=6.673\times 10^{-8}cm^{3}/gs^{2},\,\, c=2.997\times 10^{10}cm/s,\,\,$.
\par 
It is hard to find a mathematical models of anisotropic fluid spheres satisfying all the physical constraints for stellar bodies like compact stars. We have successfully accomplished to model such type of objects opting Buchdahl potential which is particularly befitting for interior geometry of compact objects and modified Chaplygin EoS which can be explicitly useful to generate models like these as well as exotic stellar objects in future endeavors.
\section*{Acknowledgments}
The authors would like to thank Council of Scientific \& Industrial Research (CSIR), India for financial support.
The authors are also thankful to Central University of Jharkhand, India for kind support.

\end{document}